\def\frac#1#2{{{{#1}}\over{{#2}}}}
\documentclass[12pt]{article}
\usepackage{amsmath}
\usepackage{wasysym}    
\usepackage{textcomp}
\usepackage{pict2e}
 \usepackage[absolute]{textpos}
\usepackage{cite}

\usepackage{xcolor, cancel, ulem}
\usepackage{graphicx}
\usepackage{hyperref}
\usepackage{pdfpages}
\usepackage{cite}

\newsavebox{\ns}
\newsavebox{\dbrane}
\newsavebox{\dbshort}

\usepackage{epsfig}
\usepackage{amssymb}

\def\appendix{{\newpage\section*{Appendix}}\let\appendix\section%
        {\setcounter{section}{0}
        \gdef\thesection{\Alph{section}}}\section}

\newcommand\ba{\begin{eqnarray}}
\newcommand\ea{\end{eqnarray}}

\definecolor{DarkGreen}{rgb}{0,.64,0}
\definecolor{gunmetal}{rgb}{0.171875, 0.207031, 0.222656}
\definecolor{chartreuse}{rgb}{.49,.98,0}
\definecolor{amethyst}{rgb}{0.59375,0.398438,0.792969}
\definecolor{brownrust}{rgb}{0.6875, 0.316406, 0.242188}
\definecolor{Violet}{rgb}{0.5,0,1}
\definecolor{BurntOrange}{rgb}{0.792969,0.332031,0}
\definecolor{FreshEggplant}{rgb}{0.59375, 0., 0.414063}
\definecolor{salmon}{rgb}{0.996094,0.507813,0.410156}
 \definecolor{FrenchRose}{rgb}{0.96875, 0.292969, 0.5625}
\definecolor{Cabaret}{rgb}{0.808594, 0.242188, 0.46875}
\definecolor{Shamrock}{rgb}{0.242188, 0.808594, 0.582031}
\definecolor{RobinsEggBlue}{rgb}{0., 0.792969, 0.792969}
\definecolor{GuardsmanRed}{rgb}{0.792969, 0., 0.}
\definecolor{Sapphire}{rgb}{0.183594, 0.328125, 0.621094}
\definecolor{Sorbus}{rgb}{0.996094, 0.429688, 0.0273438}
\definecolor{Red}{rgb}{1,0,0}
\definecolor{Blue}{rgb}{0,0,1}
\definecolor{Black}{rgb}{0,0,0}
\definecolor{Green}{rgb}{0,1,0}
\definecolor{thistle3}{rgb}{0.800781, 0.707031, 0.800781}
\definecolor{thistle4}{rgb}{0.542969, 0.480469, 0.542969}
\definecolor{DarkTurquoise}{RGB}{0,206,209}
\definecolor{turquoise4}{RGB}{0,134,139}

\definecolor{Purple}{rgb}{0.808594, 0.242188, 0.46875}
\newcommand{\purple}[1]{{\color{Purple} {#1}}}

\newcommand{\TENSOREDIT}[1]{{}}

\newcommand{\sshg}[1]{}

\newcommand{\sisg}[1]{}

\newcommand{\nn}{\nonumber}

\def\Dslash{\,\,{\raise.15ex\hbox{/}\mkern-12mu D}}
\def\Dbarslash{\,\,{\raise.15ex\hbox{/}\mkern-12mu {\bar D}}}
\def\delslash{\,\,{\raise.15ex\hbox{/}\mkern-9mu \partial}}
\def\delbarslash{\,\,{\raise.15ex\hbox{/}\mkern-9mu {\bar\partial}}}
\def\pslash{\,\,{\raise.15ex\hbox{/}\mkern-9mu p}}
\def\calDslash{\,\,{\raise.15ex\hbox{/}\mkern-12mu {\cal D}}}

\newcommand{\hh}{{1\over 2}}

\renewcommand{\ll}{_}
\newcommand{\uu}{^}
\newcommand{\pp}{\partial}

\renewcommand{\exp}[1]{{\rm exp}\{#1\}}

\newcommand{\expp}[1]{{\rm exp}( #1 ) }

\newcommand{\m}{\mu}

\renewcommand{\m}{\mu}

\newcommand{\s}{\sigma}
\renewcommand{\t}{\tau}

\newcommand{\g}{\gamma}
\renewcommand{\a}{\alpha}
\renewcommand{\r}{\rho}

\renewcommand{\o}{\omega}
\newcommand{\e}{\epsilon}

\newcommand{\sqd}{^2}

\newcommand{\pb}{{\bar{\partial}}}

\renewcommand{\hh}{{1\over 2}}
\renewcommand{\gg}{\nabla}

\newcommand{\eee}[1]{\ba{#1}\ea}

\renewcommand{\t}{\tau}

\renewcommand{\b}{\beta}

\newcommand{\llsk}{\hskip .5in}

\newcommand{\D}{\Delta}

\newcommand{\apr}{{\alpha^\prime} {}}

\newcommand{\IZ}{\relax\ifmmode\mathchoice
{\hbox{\cmss Z\kern-.4em Z}}{\hbox{\cmss Z\kern-.4em Z}}
{\lower.9pt\hbox{\cmsss Z\kern-.4em Z}} {\lower1.2pt\hbox{\cmsss
Z\kern-.4em Z}}\else{\cmss Z\kern-.4em Z}\fi} \font\cmss=cmss10
\font\cmsss=cmss10 at 7pt
\newcommand{\inbar}{\,\vrule height1.5ex width.4pt depth0pt}
\newcommand{\IC}{{\relax\hbox{$\inbar\kern-.3em{\rm C}$}}}
\newcommand{\IQ}{{\relax\hbox{$\inbar\kern-.3em{\rm Q}$}}}
\newcommand{\IP}{\relax{\rm I\kern-.18em P}}

\renewcommand{\k}[1]{{k_{#1}}}

\newcommand{\ed}{\dot{e}}

\renewcommand{\k}{\kappa}

\newcommand{\cc}{{\cal C}}

\renewcommand{\cc}{{c_1}}

\renewcommand{\o}{\omega}

\renewcommand{\cc}{c}

\newcommand{\pr}{^\prime}

\newcommand{\IR}{\relax{\rm I\kern-.18em R}}
\def\blfootnote{\xdef\@thefnmark{}\@footnotetext}

\renewcommand{\cc}[1]{\cite{#1}}

\newcommand{\bm}{\begin{matrix}}
\renewcommand{\em}{\end{matrix}}

\newcommand{\ee}[1]{\ba {#1} \ea}

\def\kst{\Xi}

\newcommand{\upp}[1]{^{({#1})}{}}

\newcommand{\phh}{{\hat{p}}}

\newcommand{\co}{{\cal O}}

\newcommand{\rr}[1]{(\ref{{#1}})}
\newcommand{\bbb}{\begin{eqnarray}}
\renewcommand{\eee}{\end{eqnarray}}
\newcommand{\een}[1]{\label{#1}\end{eqnarray}}

\newcommand{\heading}[1]{\vspace{6mm}\begin{center}\it #1 \rm \end{center}}

\def\bi{\begin{itemize}}
\def\ei{\end{itemize}}

\def\ed{\end{document}}

\def\cc{{\cal C}}

\renewcommand{\rr}[1]{(\ref{#1})}

\def\cc{\,}

\def\r{\rho}

\def\cc{\,}

\def\Zb{{\bar{Z}}}

\newcommand{\lp}[1]{_{({#1})}}
\newcommand{\lprm}[1]{_{({\rm {#1}})}}

\newcommand{\uprm}[1]{^{({\rm {#1}})}}

\usepackage{fancyhdr}
\renewcommand{\eqref}[1]{\rr{#1}}
\usepackage{dsfont}

\def\binv{\blue{{\cal B}}}

\def\outt#1{}

\newcommand{\redd}[1]{{\color{Red} {#1}}}

\definecolor{Purple}{rgb}{0.808594, 0.242188, 0.46875}

\renewcommand{\bm}{\begin{matrix}}
\renewcommand{\em}{\end{matrix}}

\newcommand{\aaa}[1]{}

\renewcommand{\caption}[1]{\begin{center}{{#1}}\end{center}}

\def\be{\begin{eqnarray}}
\def\ee{\end{eqnarray}}
\def\nn{\nonumber}

\newcommand{\WhiteOutWithNotification}[1]{ {  \color{Red}  (  A SECTION HAS BEEN WHITED OUT HERE. \cc   )  } }
\newcommand{\WhiteOut}[1]{}
\newcommand{\RoundIEdit}[1]{}
\newcommand{\RoundHEdit}[1]{}
\newcommand{\RoundGEdit}[1]{}
\newcommand{\RoundFEdit}[1]{}
\newcommand{\RoundEEdit}[1]{}
\newcommand{\EMVertEdit}[1]{}
\newcommand{\HugeEditB}[1]{}
\newcommand{\HugeEditA}[1]{}
\newcommand{\RoundDEdit}[1]{}
\newcommand{\RoundCEdit}[1]{}
\newcommand{\RoundBEdit}[1]{}
\newcommand{\RoundAEdit}[1]{}

\newcommand{\blue}[1]{{\color{Blue}{#1}}}

\def\Zb{{\bar{Z}}}

\def\redlowdash{{\color{Red}{\rule[-0.5ex]{2pt}{0.4pt}}}}
\def\redmiddash{{\color{Red}{\rule[+0.5ex]{2pt}{0.4pt}}}}

\def\cute{{\lower3.5pt\hbox{\sixly
  \kern-.21pt \char58 \kern-.21pt }}}

\def\midcute{{\lower-1.0pt\hbox{\sixly
  \kern-.21pt \char58 \kern-.21pt }}}

  \def\lowcute{{\lower3.5pt\hbox{\sixly
  \kern-.21pt \char58 \kern-.21pt }}}
  
  \def\redmidcute{{\color{Red} \midcute}}
  
  \def\redlowcute{{\color{Red} \lowcute}}
   
    \def\bluelowcute{{\color{Blue} \lowcute}}

   \def\swave{\bgroup \markoverwith \midcute \ULon} 
  
  \def\redswave{\bgroup \markoverwith \redmidcute \ULon} 
  
  \def\reduline{\bgroup \markoverwith \redlowdash \ULon}
  
   \def\blueuline{\bgroup \markoverwith \bluelowdash \ULon}
  
   \def\reduwave{\bgroup \markoverwith \redlowcute \ULon}
   
   \def\blueuwave{\bgroup \markoverwith \bluelowcute \ULon}
  
  \def\redsout{\bgroup \markoverwith \redmiddash \ULon}
  
   \def\bluesout{\bgroup \markoverwith \bluemiddash \ULon}
   
   \def\Irrel{\bgroup \markoverwith {{\color{Red} {\bf X}}} \ULon}

\newcommand{\eqirrel}[1]{\rdots}
  
\let\oldcancel\cancel
\renewcommand\cancel[1][black]{%
  \def\CancelColor{\color{#1}}%
  \oldcancel}

   \let\oldbcancel\bcancel
\renewcommand\bcancel[1][black]{%
  \def\CancelColor{\color{#1}}%
  \oldbcancel}

\def\bca{\begin{cases}}
\def\eca{\end{cases}}

\def\rdots{\redd{\circ\circ\circ}}

\renewcommand{\upp}[1]{^{({#1})}}

\def\lsim{\mathrel{\lower0.3em\hbox{$\stackrel{\textstyle <}{\sim}$}}}
\def\gsim{\mathrel{\lower0.3em\hbox{$\stackrel{\textstyle >}{\sim}$}}}

\def\negspace{\kern -0.4em}

\def\dvec{\raise 0.3 em \hbox{$^\leftrightarrow$} \kern -0.77 em}

\def\omegahat{\hat%
	{\setbox0=\hbox{$\omega$}%
		\kern-.025em\copy0\kern-\wd0
		\kern.05em\copy0\kern-\wd0
		\kern-.025em\raise.0433em\box0}}

\def\phh{{\blue{\hat{\phi}}}}

\def\pol#1{}

\def\inv{X}
\def\inv{\blue{\tt{Inv}}}
\def\inv{{\blue{{\cal I}}}}

\def\phc{{\purple{\varphi}}}

\def\SilentOthersComment#1{{}}

\newcommand{\defout}[1]{}

\def\tha#1{\breve{#1}}

\def\dha#1{\hat{\hat{{#1}}}}
\def\blat{\dha{\phi}}
\def\plat{\dha{\Phi}}

\begin{document}

\begin{titlepage}
\begin{flushright}
IPMU16-0117\\
CALT-TH-2016-025\\
\end{flushright}
\vspace{8 mm}
\begin{center}
  {\large \bf Boundary operators \\ \vspace{2mm} in effective string theory}
\end{center}
\vspace{2 mm}
\begin{center}
{Simeon Hellerman$^1$, 
Ian Swanson}\\
\vspace{6mm}
{\it $^1$Kavli Institute for the Physics and Mathematics of the Universe\\
The University of Tokyo \\
 Kashiwa, Chiba  277-8582, Japan\\}
\vspace{6mm}

\end{center}
\vspace{-4 mm}
\begin{center}
{\large Abstract}
\end{center}
\noindent
Various universal features 
of relativistic rotating strings
depend on the organization of allowed local operators on the
worldsheet.  In this paper, we study the set of 
Neumann boundary operators in effective string theory, 
which are relevant for the controlled 
study of open relativistic strings with freely moving endpoints.  
Relativistic open strings are thought to encode the dynamics of confined quark-antiquark 
pairs in gauge theories in the planar approximation. 
Neumann boundary operators can be organized by their behavior under scaling of the target space coordinates $X\uu\m$, and the set of allowed $X$-scaling exponents is bounded above by $+1/2$ and unbounded below.  
Negative contributions to $X$-scalings come from powers of a single invariant, or
``dressing" operator, which is bilinear in the embedding coordinates. 
In particular, we show that all Neumann boundary operators are dressed 
by quarter-integer powers of this invariant, and we demonstrate how this rule arises from various ways of 
regulating the short-distance singularities of the effective theory.
\vspace{1cm}
\begin{flushleft}
\today
\end{flushleft}
\end{titlepage}

\tableofcontents
\newpage
\numberwithin{equation}{section}
\section{Introduction} 
For many purposes, it is useful to study the dynamics of relativistic
strings in the framework of a Wilsonian effective
field theory on the string worldsheet.  This enterprise is known as effective string theory.
A manifestly Poincar\'e-invariant version of this theory was invented 
in \cite{Polchinski:1991ax}.  This theory was more recently elucidated
by embedding it into the Polyakov formalism
\cite{HariDass:2007gn,Hellerman:2014cba}, which simplifies the
construction of vertex operators and the renormalization of
short-distance singularities of the worldsheet path integral.  
This simplified covariant formalism has
been applied to calculate subleading terms in the perturbative 
expansion of the mass-squared of rotating string states at large 
angular momentum $J$ \cite{PRL}.  

Open relativistic strings with 
freely moving endpoints are of particular interest, 
as these objects are believed to describe the 
dynamics of confined quark-antiquark pairs in gauge theories in the
planar approximation.  In the covariant formalism, the condition of 
freely moving endpoints corresponds to a Neumann boundary condition 
on the embedding coordinates $X\uu\m$, at leading order in the
expansion around large angular momentum $J$.  To analyze
higher-order corrections and renormalize such theories correctly, it is important to
characterize the spectrum of boundary operators in the open 
worldsheet theory with Neumann boundary conditions, as has been 
done \cite{Aharony:2013ipa, AharonyUnPub} for operators in the
interior of the worldsheet and for boundary operators with Dirichlet 
boundary conditions.

In all cases, local operators are organized hierarchically under 
an assignment of $X$-scaling dimension, which encodes the scaling 
dimension of the term in the standard Wilsonian sense.  (We use 
the term ``$X$-scaling dimension'' to distinguish the
$X$-scaling from the scaling dimensions 
of operators under the residual Virasoro symmetry of the
Polyakov formalism after gauge-fixing the
metric. The Virasoro 
algebra is a residual gauge symmetry of the Polyakov action, and 
the weights of all physical states and operators under it are
determined by gauge symmetry.)
The spectrum of 
$X$-scaling dimensions of operators depends on the details of 
the particular effective string theory under consideration, but in 
all cases is bounded above, and continues discretely downwards towards
$-\infty$.  Negative $X$-scalings come from negative powers of the ``dressing" operator, a
distinguished bilinear invariant
of $X$, which compensates the conformal scaling dimension of a
nonsingular numerator. The numerators are polynomials in derivatives of $X$.

For bulk operators and Dirichlet boundary operators, the operator 
spectrum is organized according to a rule for dressing operators 
for each case, specifying the unique operator that can occur to
negative or fractional powers, and which fractional exponents can
occur.  In the case of bulk operators in conformal gauge, the dressing 
rule amounts to the condition that operators are dressed with negative integer powers
of the dressing operator $\inv\ll{11}$, where\footnote{Our worldsheet coordinate conventions are 
$\s^\pm \equiv \s^0 \pm \s^1 = \tau \pm \s$, $\pp_\pm = \frac{1}{2}(\pp_0 \pm \pp_1)$.}
\bbb
\inv\ll{pq}\equiv \pp\ll +\uu p X \cdot \pp\ll - \uu q X\ .
\een{IpqDef}

For boundaries where some coordinates $X$ have
Neumann boundary conditions and some have Dirichlet, the dressing
rule for boundary 
operators is that the operator dressing consists
only of half-integer powers of ${\binv}\lp{11} \equiv \dot{X}_{\textsc{Neumann}}\sqd$,
where we introduce the notation
\bbb
{\binv}\lp{pq} \equiv \pp\ll{\s\uu 0}\uu p X \cdot  \pp\ll{\s\uu 0}\uu q X
\een{BpqDef}
for boundary operators.
In each case there exist a number of simple motivations for the
dressing rule, which we shall discuss briefly in Section
\ref{IntuitiveDerivations}.  For instance, the correct result for the dressing rule 
can be stated in terms of a pure power principle:  The dressing
comes in negative powers of the most relevant bilinear invariant operator.
In the example above, the dressing consists of powers of the operator ${\cal O}=\binv\lp{11}^{1/ 2}$.

The dressing rule in the case of Neumann boundary conditions is 
equally simple but less familiar.  We will show below that at Neumann 
boundaries, operators are dressed with negative quarter-integer powers
of the operator ${\binv}\lp{22}$, 
or, equivalently, negative integer powers of the
operator ${\cal O}\ll{{({\rm quark})}} \equiv {\binv}\lp{22}\uu{{1/ 4}}$,
which encodes the leading physical effect of an infinitesimal change in the mass of the quark.
This is another instance of 
the pure power principle: The
operator ${\cal O}\ll{{({\rm quark})}}$ is the invariant with the 
largest $X$-scaling in the theory with Neumann boundaries.

The dressing rule for Neumann boundary operators is particularly 
physically significant, because there is an anomaly term in the 
Lagrangian density with a universal coefficient\footnote{We are using the
term ``universal" according to its usual meaning in the context
of effective field theory, i.e., that the value
of the asymptotic intercept does not depend on the values of adjustable parameters in the effective worldsheet action.
If one considers a universality class with different low-energy degrees of freedom or symmetries, the value of 
the asymptotic intercept may of course differ.}
that has a non-integrable divergence near a Neumann boundary.  This singularity
does not signify a breakdown of the effective theory. Rather, the singularity 
is removed by a familiar procedure of short-distance regularization, and the divergence 
in the quantum effective action when the regulator is removed 
must be cancelled by a boundary counterterm. In this case the 
counterterm is the quark 
mass operator ${\cal O}\ll{{({\rm quark})}}$ itself
\cite{PRL}.  Thus, demonstrating the 
renormalizability of the effective string theory with Neumann 
boundary conditions depends on the use of the correct dressing rule.

Moreover, the universality of the large-$J$ spectrum (and other observables) 
at relative order $J\uu{-1}$ also depends on the use of the correct 
dressing rule: There are neither bulk nor boundary terms in the action
with adjustable coefficients at order $|X|\uu 0$ (which is 
order $|X|\uu{-2}$ relative to the Nambu-Goto action). 
Amplitudes are therefore universal at relative order $J\uu{-1}$, since the
length of a string scales with its angular momentum as $({\rm length}) \sim J\uu{\hh}$.  
If there were such an operator at  order $|X|\uu 0$, amplitudes would
not be universal at relative order $J\uu{-1}$.  
Demonstrating the absence of order-$|X|\uu 0$ 
operators at Neumann boundaries, and therefore the
universality of the asymptotic intercept of the Regge trajectory, thus depends on 
the correct derivation and application of the dressing rule. 

The goal of the paper is to explain the origin of the dressing rule for 
Neumann boundary operators.  To provide some context, and to guide the
subsequent analysis, we will begin in the following section by motivating the 
rule on heuristic grounds.  In Section \ref{Liouville} we will provide a more concrete 
derivation of the dressing rule starting from an ultraviolet-complete 
worldsheet theory.  As a demonstration that the conclusions are
in fact independent of the details of any particular UV completion of
the effective string theory, we take yet another approach in Section
\ref{FiniteSepReg}.  There, we adopt a displaced-boundary UV regulator scheme and
show how the dressing rule emerges from this regulator for a few specific operators of interest.
We conclude in Section \ref{conclusions} with a broader discussion of
some of the physical consequences of the dressing rule.
As supplementary material to both the study of boundary operators in
effective string theory, and to the larger goal of understanding
certain aspects of the strong-coupling dynamics of QCD via string
theory, we provide in the appendices a more detailed and fully
gauge-invariant calculation of the asymptotic Regge intercept (a
result originally presented in \cite{PRL}), 
as well as other calculational details that support the
conclusions drawn in the main body of the paper.

\section{Heuristic motivation}\label{IntuitiveDerivations}
We have proposed a boundary operator dressing rule in effective string
theory, which states that at Neumann 
boundaries, operators are dressed with negative quarter-integer powers
of the operator ${\binv}\lp{22}$.  In this section we will motivate
this claim on heuristic grounds,
based on minimality, naturalness and on the pure power principle discussed above.
Following this, we will turn to a more detailed derivation of the Neumann-boundary
dressing rule.

\subsection{Dressing rules for bulk and Dirichlet boundary operators}
\label{bulkAndDirichlet}
\heading{Minimality}
We begin with an overview of bulk worldsheet operators and Dirichlet boundary operators.
In ref.~\cite{Polchinski:1991ax}, the dressing rule for bulk operators was assumed rather than
derived.  There, Polchinski and Strominger (PS) introduced a singular
interaction term in the Lagrangian of order $|X|\uu 0$ for spacetime
dimension $D\neq 26$:
\be
{\cal L}_{\rm PS} = \frac{\b}{2\pi} \frac{\inv\ll{12}\inv\ll{21}}{\inv\ll{11}^2}\ ,
\label{psterm}
\ee
which compensates for the conformal anomaly by contributing $\Delta c
= 26 - D$ to the central charge of the conformal dynamics of the
embedding coordinates $X^\mu$.  
Rather than basing the form of this term explicitly on general restrictions
descending from the symmetries and constraints of the theory,
the authors arrived at eqn.~\rr{psterm} based on the specific
requirement that the conformal anomaly cancel for
small perturbations around the static string solution.  
The denominator of the singular term is $\inv\ll{11}^2$, from
which it follows that integer powers of $\inv\ll{11}$ must be allowed in the bulk operator
dressing rule.  Singularities in operator products between powers of the anomaly term
above have denominators that are integer powers of $\inv\ll{11}$ only.  To all orders in
large-$|X|$ perturbation theory, then, the rule that bulk operators appear
dressed strictly with integer powers of $\inv\ll{11}$ is stable against quantum corrections.
An $\inv\ll{11}$ dressing rule can thus be taken as the minimal rule that is consistent with anomaly
cancellation and stable against quantum corrections.
It is this notion of minimality that we will propose be applied to the
case of Neumann boundaries.

\heading{Naturalness and genericity: The pure power principle}
For bulk operators and Dirichlet boundary operators, the dressing
operator is the unique monomial with the lowest $X$-scaling obeying the
required symmetries, and it is automatic that such an operator is a conformal 
tensor.  In other words, the dressing rule for bulk and Dirichlet operators 
takes the form of negative integer powers of the most relevant operator 
that can be expressed as a power of a bilinear invariant.
Below we will promote this structure to an ansatz for 
dressing operators in general, and apply it to the case of 
Neumann boundaries.  

As discussed briefly in \cite{Hellerman:2014cba}, the bulk dressing rule 
follows in some sense from naturalness.  If an 
operator 
of nonmarginal conformal weight is 
dressed to conformality with some other operator, 
and we assume the latter is generic (in the sense 
of being a linear combination of all possible operators 
of the appropriate conformal weight), then at large $J$ 
the dressing operator will be dominated by powers of 
a single operator, $\inv\ll{11}$ (that is, 
the operator that contains the most powers of $X$ per conformal dimension).
The same conclusion holds for 
Dirichlet boundaries:  Naturalness dictates that the 
dressing operator be dominated by powers of ${\binv}\lp{11}$.  
We will generalize this notion of naturalness 
to the case of Neumann boundaries as well.

\subsection{The Neumann dressing rule}
Let us now proceed with a proposal for an operator dressing rule at Neumann boundaries that
satisfies the same properties as the bulk and Dirichlet dressing rules summarized above.
To motivate the claim on heuristic grounds, 
we begin with the ansatz that our dressing operator be a monomial in bilinears in derivatives of the
embedding coordinates $X$.  To satisfy the principle of
naturalness, we require the monomial to have the highest possible $X$-scaling for its worldsheet conformal
dimension (or, equivalently, the lowest possible worldsheet conformal
dimension for its $X$-scaling).  
Both $X$-scaling and (in the semiclassical regime) 
conformal dimension are additive under multiplication, so the ratio
of $X$-scaling to worldsheet conformal dimension is maximized by
powers of a single bilinear.  Therefore, we will identify the nonvanishing bilinear with lowest possible worldsheet conformal 
scaling dimension and see that it is unique up to operator equivalences.

\newcommand{\bdol}[1]{${#1}$}
First, one can use the equations of motion to reduce all derivatives of
\bdol{X} to the form \bdol{\pp\ll 0\uu p X} 
or \bdol{\pp\ll 0\uu p \pp\ll 1 X}, where Neumann boundary conditions 
can be used to eliminate the latter.\footnote{Where we now replace explicit worldsheet directions
$\sigma^{0,1}$ with $(0,1)$ indices.}
The only invariant candidate operators remaining are then of the form
\be
{\binv}\lp{pq} = \pp\ll 0 \uu p X \cdot \pp\ll 0 \uu q X\ ,
\ee
introduced in \rr{BpqDef}.  It is important 
to note here that when we eliminate other possible operators from this search, we are only
doing so modulo operators of lower $X$-scaling.  For the purpose 
of estimating leading-order scalings and proceeding order-by-order in $J$,
this is always sufficient, and whenever we refer to eliminating operators 
by appealing to constraints or to the equations of motion, we will always mean it in this sense.

The operator ${\binv}\lp{11}$ is proportional to the leading-order stress tensor, and thus vanishes 
as an operator, modulo operators of lower $X$-scaling.  The operator ${\binv}\lp{12}$ is a tangential derivative of
${\binv}\lp{11}$, and also vanishes, modulo operators of lower $X$-scaling.   
Next we come to the operators ${\binv}\lp{22}$ and ${\binv}\lp{13}$.  The sum of these 
is proportional to a second tangential derivative of ${\binv}\lp{11}$, and
we can eliminate ${\binv}\lp{13}$ as an operator in favor of ${\binv}\lp{22}$, modulo operators of lower $X$-scaling. 

The operator ${\binv}\lp{22}$ itself is an independent, gauge-invariant operator.  For instance, it takes a nonzero
expectation value, proportional to $J$, in the lowest state of angular momentum $J$ that satisfies the Virasoro constraints.
It is natural to propose, then, that ${\binv}\lp{22}$ should be the dressing operator for effective string theories with
Neumann boundaries.  More precisely, and following the pure-power principle, the proposed dressing rule states
that the basis for symmetry-preserving
boundary operators in conformal gauge, with Neumann boundary conditions, is
\bbb
\co \in {\rm Span} \bigg \{ \binv\lp{22}\uu {-{1\over 4} \left( \D - 1\right)} \cdot \prod\ll i \binv\lp{p\ll i q\ll i}\ ;
\qquad \D \equiv \sum\ll i p\ll i + q\ll i   \bigg \} \ .
\een{DressingRuleStatement}

Before moving on to a more principled derivation of this rule, we pause to
make some final comments about the proposed $\binv\lp{22}$ dressing rule for Neumann boundaries:
\bi
\item{Ordinarily, we expect that perturbations of the Lagrangian in conformal gauge must be Virasoro primaries of
weight one.  The operator $\binv\lp{22}\uu a$ is not quite Virasoro primary; the lowering operator $L\ll 1$ does not annihilate it identically,
but gives an operator proportional to $\binv\lp{12} \cc \binv\lp{22}\uu{a-1}$, which is a total derivative.  The theory is physically invariant under a 
change of conformal frame because conformal transformations
of the perturbation vanish identically after integration
over the boundary.  Note that this could not happen
in a unitary theory, as the total derivative would act on an operator
of weight negative one.  }

\item{The rule has direct phenomenological consequences: The leading gauge-invariant boundary operator
${\binv}\lp{22}$ has been studied in other gauges \cite{Baker:2002km,Wilczek:2004im}, 
and the insertion of its fourth root into the boundary action corresponds to the leading effect of an infinitesimal variation of the quark mass at large $J$.}
\item{Finally, in actual UV-complete worldsheet theories, where the non-Goldstone degrees of freedom 
can be integrated out, the dressing rule for Neumann boundaries in the effective theory is indeed 
the one we have described. In what follows, we will demonstrate this
in full detail in one such UV completion of the effective theory,
taking the form of a perturbed Liouville theory \cite{Hellerman:2014cba}.  We will also comment 
in more detail on how this rule satisfies the minimality principle outlined above.}
\ei

\section{Effective strings from perturbed Liouville theory}\label{Liouville}
We have now motivated the dressing rule for Neumann boundaries on heuristic grounds.  Namely, 
Neumann boundary operators are dressed with negative quarter-integer powers
of the operator ${\binv}\lp{22}$.  In this section we
derive this statement rigorously from the starting point of a particular UV complete worldsheet theory.

In \cite{Polchinski:1991ax}, Polchinski and Strominger present a
microscopic model from which an effective string description might emerge, in terms
of a $(D+1)$-dimensional string theory with a Liouville direction.  
At large $D$, the PS-Liouville Lagrangian looks like
\be
{\cal L} = \frac{|D|}{24\pi} (\pp\phi)^2 + \m^2e^{-2\phi} + \mu'^{-2}
e^{2\phi} \inv\ll{11}^2  + ({\rm ~ \phi-independent~}) \ .
\label{PSL}
\ee
Here, $\m$ and $\m\pr$ are arbitrary mass
parameters.  One can introduce open strings into this model by introducing a
space-filling brane into the Liouville theory.

\subsection{Origin of the dressing rule for bulk operators} 
We begin our analysis of this theory by extracting the form of the 
dressing rule for bulk operators noted above.  The intent is that this will provide 
context for the subsequent analysis in the boundary theory.  
In particular, we derive the $\inv\ll{11}$ dressing
rule for the closed string theory, motivated on general grounds in
Section \ref{bulkAndDirichlet} above.  
We will derive this dressing rule in a class of UV-complete models that generalizes the
construction of ref.~\cite{Polchinski:1991ax} to a considerable extent, 
showing that the dressing rule emerges for completely generic gauge-invariant and 
Poincar\'e-invariant operator perturbations of the theory.
The ideas in this subsection are to some extent implicit in
\cite{Polchinski:1991ax}; we make them explicit to show that
they indeed generalize to the case when open strings are included.

First, we redefine the Liouville field in such a way that the second
and third terms in the equations of motion scale equally, and the
kinetic term for $\phi$ can be neglected in the large-$X$ regime:
\be
e^\phi & = & \sqrt{\m\cc\m\pr} \cc \inv\ll{11}^{-\hh} e^{\hat\phi} \ ,  \nn\\
\phi & = & \hat{\phi} - \hh \cc \log \left( {{\inv\ll{11}}\over{\m\cc\m\pr}} \right) \ .
\label{BulkLiouvShift}
\ee
Note that this field transformation becomes singular
whenever $\m$ or $\m\pr$ vanish.  This redefinition
is specifically adapted to a situation where $\phi$ gets
a minimum of its effective potential for a long string, which
happens only when both $\m$ and $\m\pr$ are nonzero. 

Under this field redefinition, the Lagrangian above becomes
\be
{\cal L} = {\cal L}\ll{|X|\uu 2} + {\cal L}\ll{|X|\uu 0} + O(|X|\uu{-2})\ ,
\ee
with
\be
{\cal L}\ll{|X|\uu 2} &\equiv& {\m\over{\m\pr}} \cc \bigg ( \expp{2\hat{\phi}} + \expp{-2\hat{\phi}} \cc \bigg ) \cc  \inv\ll{11} 
+ \biggl ( {\rm terms~subleading~in~D}\biggl )
\nn\\
&=& {{2\m}\over{\m\pr}} \cc  \inv\ll{11} 
+ \biggl ( {\rm terms~involving~} \hat{\phi} {\rm ~fluctuations} \biggl ) 
+ \biggl ( {\rm terms~subleading~in~D}\biggl )\ ,
\nn\\
{\cal L}\ll{|X|\uu 0} &\equiv& {\cal L}\ll{\rm PS} +  \biggl ( {\rm terms~involving~} \hat{\phi} {\rm ~fluctuations} \biggl ) 
+ \biggl ( {\rm terms~subleading~in~D}\biggl )\ .
\nn\\
&&
\ee
In particular, the order $|X|\uu 0$ term agrees with 
the PS anomaly Lagrangian up to terms of order $|D|\uu 0$, as pointed out
in \cite{Polchinski:1991ax}.  

We observe that the effective theory in \cite{Polchinski:1991ax} can be derived from a much larger class of
microscopic models described by perturbations of the $(D+1)$-dimensional string theory with one Liouville
direction.  Adding higher-derivative terms dressed to conformality with Liouville exponentials leads
to the same scaling for $\hat{\phi}$ and the same
coefficient for the anomaly term in the effective Lagrangian at
leading order.  For instance, consider a more general Lagrangian of the form
\be
{\cal L} &=& \frac{|D|}{24\pi} (\pp\phi)^2 + \m^2e^{-2\phi} +  \sum\ll{q \geq 0} a\ll q \cc \m\uu{-2q} \expp{2q\phi}
\cc \inv\ll{11}^{q+1}
\nn\\
&=& \frac{|D|}{24\pi} (\pp\phi)^2 + \inv\ll{11} \cc F(y) \ ,
\label{GeneralX11Model}
\ee
with
\be 
y \equiv \m\uu{-2} \cc \inv\ll{11} \cc \expp{2\phi}\ ,
\ee
where $a\ll 0$ is the microscopic string tension (i.e., the string
tension in the full $(D+1)$-dimensional string theory), and
$a\ll 1, a\ll 2, \cdots$ are the values of massive stringy
condensates.  If $(y\ll 0, F(y\ll 0))$ is the global minimum of
$F(y)$, then we can shift the Liouville field to $\hat{\phi}$, where
\be
e^\phi  &=&  \m\cc \sqrt{y\ll 0}  \cc \inv\ll{11}^{-\hh} e^{\hat\phi}\ ,
\nn\\
\phi  &=&  \hat{\phi} - \hh \cc \log \bigg ( {{\inv\ll{11}}\over{\m\sqd\cc y\ll 0 }} \bigg ) \ ,
\label{phiShift}
\ee
whereby we obtain
\bbb
{\cal L} &=& {\cal L}\ll{|X|\uu 2} + {\cal L}\ll{|X|\uu 0} +
O(|X|\uu{-2})\ ,
\nn\\
{\cal L}\ll{|X|\uu 2} &\equiv& F(y\ll 0 ) \cc \inv\ll{11} 
+ \biggl ( {\rm terms~involving~} \hat{\phi} {\rm ~fluctuations} \biggl ) 
+ \biggl ( {\rm terms~subleading~in~D}\biggl )
\ ,
\nn\\
{\cal L}\ll{|X|\uu 0} &\equiv& {\cal L}\ll{\rm PS} +  \biggl ( {\rm terms~involving~} \hat{\phi} {\rm ~fluctuations} \biggl ) 
+ \biggl ( {\rm terms~subleading~in~D}\biggl ) \ .
\nn\\
\eee
Here, the global minimum $F(y\ll 0)$ assumes the role of the effective string tension in the $D$-dimensional
effective string theory.
The coefficient of the PS anomaly Lagrangian is, as expected, independent of the form of
$F(y)$ and in agreement, at leading order in $|D|$, with the value required 
\cite{Polchinski:1991ax} to compensate the $O(|D|)$ contribution to the central charge deficit.

The terms linear in $\hat{\phi}$ vanish at orders $|X|\sqd$ and
$|X|\uu 1$; as with \rr{BulkLiouvShift}, we have chosen the shift in $\phi$ \rr{phiShift} so that $\hat{\phi} = 0$ is a
solution to the classical equations of motion for $\phi$ with the
$O(|X|\uu 0)$ kinetic terms omitted.  The mass-squared of the
$\hat{\phi}$ fluctuation is of order $|X|\sqd$, and it can be integrated out.  The leading
contribution of the path integral over $\hat{\phi}$ is of order $|D|\uu 0 \cc |X|\uu 0$, which 
acts only to shift the coefficient of the Polchinski-Strominger anomaly term by one unit of central charge.

We can now perturb the microscopic string theory with arbitrary gauge-invariant operators
and investigate the structure of resulting perturbations of the effective theory.  It is immediately clear that the
bulk dressing rule is respected.  Perturbations of the microscopic theory are generated by monomials in derivatives
of $X$ and $\phi$, dressed with Liouville exponentials.  But, as we have seen, the classical elimination of the Liouville
degree of freedom $\phi$ is such that all Liouville exponentials become powers of $\inv\ll{11}$ in the effective theory.

Now let us consider several sources of error whose discussion we have omitted in the above derivation of the
bulk dressing rule; the corresponding corrections will be seen to respect the same dressing rule as the leading
terms, in both the large-$D$ and large-$X$ sense.

\heading{Corrections to the classical solution for $\phi$}
In the above discussion, we shifted $\phi$ so that $\hat{\phi} = 0$ is a minimum to accuracy
up to and including order $|X|\uu{-1}$.  
We can study the effects of the kinetic term order by order in a large-$|X|$ expansion,
however.  Expanding the classical solution $\phi\ll *$ as $\phi\ll *\upp 0 + \phi\ll * \upp 1 + \cdots$, where
$\phi\ll *\upp n$ is of order $|X|\uu{-2n}$, one can solve for each
order iteratively.  
For instance, in the case
of the original PS-Liouville model \cite{Polchinski:1991ax}, 
we have\footnote{Here, $\pp$ and $\pb$ are the 
usual derivatives with respect to the holomorphic 
conformal coordinates on the worldsheet.}
\bbb
\phi\ll *\upp 1 = {{|D|}\over{24\pi}} \cc {{\m\pr}\over{4\cc \m }} \cc {{\pp\pb \phi\ll * \upp 0}\over{\inv\ll{11}}} \ .
\eee
This correction obviously satisfies the dressing rule.  More generally, it is straightforward to show inductively
that the $n^{\rm th}$ correction
to the classical solution is always of the form
\bbb
\phi\ll * \upp n =  P\ll n + {{|D|}\over{24\pi}} \cc {{\m\pr}\over{4\cc \m }} \cc {{\pp\pb \phi\ll * \upp {n-1}}\over{\inv\ll{11}}} \ ,
\eee
where $P\ll n$ is a polynomial in $\phi\ll *\upp 1$ through $\phi\ll *\upp{n-1}$, whose total $X$-scaling is exactly
$|X|\uu{-2n}$.  (The polynomials $P\ll n$ come from the large-$|X|$ expansion of the exponentials
$\exp{\pm 2 \cc (\phi\ll *\upp 1 + \phi\ll *\upp 2 + \cdots)}$.)
Thus, the dressing rule holds to all orders in the large-$|X|$ expansion of the classical solution
at large $D$.

In the above, we have used leading-order, large-$D$ expressions for the classical action.  At finite $D$, we replace $|D|$ with $26-D$ and 
supplement $\inv\ll{11}\sqd$ with subleading $D$-dependent terms,
proportional to $D\uu{-1} \cc (\pp\ll - X)\sqd (\pp\ll + X)\sqd$, in
the form of the irrelevant perturbation.  Such terms are required to make
the perturbation a Virasoro primary of weight one (in particular, see eqn.~(20) of \cite{Polchinski:1991ax}).

\heading{Subleading large-$D$ corrections from quantum effects}
The large-$D$ regime suppresses quantum corrections to the classical elimination of
the $\hat{\phi}$ fluctuations, in terms of contributions to the
effective action for the $X$ fields.  
We can also consider corrections to the Wilsonian action for $X$ in perturbation theory when we integrate out
$\hat{\phi}$ at one or more loops.  The resulting effective action is
of course complicated, but we emphasize that all Feynman
diagrams correcting the classical effective action for $X$ give terms obeying the
bulk dressing rule.  

To see this, let $M\sqd\ll{\hat{\phi}}$ be the tree-level mass of the
$\hat{\phi}$-fluctuation, while $\{
C\upp{3,4,5,\cdots}\ll{\hat{\phi}}\} $ denote, collectively, 
its cubic, quartic, quintic, etc., self-couplings.  The general structure of the effective action 
at a given order of perturbation theory will always have elements of $\{ C\upp p\ll{\hat{\phi}}\}$ in the numerator, and
powers of $M\ll{\hat{\phi}}$ in the denominator.  

In models \rr{PSL} or \rr{GeneralX11Model}, the mass of the $\hat{\phi}$ fluctuation is exactly proportional
to $\inv\ll{11}$, and so the form of the effective action, order by order in perturbation theory, is given
by polynomials in $\inv\ll{pq}$, dressed with negative powers of $\inv\ll{11}$. In $1/D$ perturbation theory, singular operators in the effective action
can come only from $\hat\phi$-propagators, since the interaction vertices for $\hat\phi$ have only positive powers of $X$ in the UV theory.
Beyond perturbation theory, the singular $X$-dependence of operators come entirely from the mass scale at which
new degrees of freedom enter, namely $M\sqd\ll\phh \propto
\inv\ll{11}$, plus terms subleading in $|X|$.  
We therefore infer that the bulk dressing rule holds for all $1/D$
quantum corrections to the effective string action as well, away from loci on the worldsheet where $\inv\ll{11}$ vanishes.

\subsection{Demonstration of the boundary dressing rule}
We now turn to the boundary theory to demonstrate explicitly the form
of the dressing rule for boundary operators in effective string
theory, for the Polchinski-Strominger deformed
Liouville theory.  We do this in the most direct possible way, expanding
around a nonsingular classical solution for the Liouville
field $\phi$ and integrating out massive fluctuations.
The first step is thus to understand how the classical solution
for the Liouville field scales in the near-boundary
region.  

We start by expanding $\inv\ll{11}$ near the boundary:
\be
\inv\ll{11} = -\hh \binv\lp{22} \s_1^2  + O(\s_1^3) \ .
\label{GeneralI11Expansion}
\ee
The expansion contains higher terms of the form $\binv\lp{pq}\s\uu{p + q - 2}$, with $\binv\lp{pq}$ defined in eqn.~\rr{BpqDef}. 
Such terms are obtained by Taylor expanding $\inv\ll{11}$ near the boundary and using the free EOM and free Virasoro constraints.  There are also terms coming from corrections to the free-field EOM due to the interactions with the Liouville field.
In this section we will estimate the $J$-scaling of such corrections and the effective terms they generate after the elimination of the 
Liouville field. 

Order by order in $\s$, equation \rr{GeneralI11Expansion} is an operator statement in the low
energy Hilbert space, rather than just a property of a particular
classical solution or matrix element in a given state.  We now
pause to emphasize this distinction.

In effective
string theory, the degrees of freedom are small fluctuations around a lowest-energy classical
solution carrying certain conserved global quantum numbers.  The lowest-lying classical solution
carrying a given set of charges is always automatically Virasoro-primary and preserves a `helical' symmetry, 
i.e.,~an invariance under a combined time translation and a global symmetry transformation, 
which in this context is a rotation of some of the target-space coordinates.
(For more details on the helical solution, see eqn.~(10) of \cite{PRL}, or Appendix \ref{derivationOfReggeIntercept}.)
In the helical solution, the values of the invariants $\inv\ll{pq}$ and $\binv\lp{pq}$ are all time-independent,
and equation \rr{GeneralI11Expansion} is simply an identity between time-independent expectation values.
However, the existence of the operator expansion \rr{GeneralI11Expansion} does not depend on
the helical property: The expansion in fact holds true even for general time-dependent
perturbations with energies of $O(1)$ above the large-$J$ ground state.
For the sake of brevity we are not explicitly indicating any
time dependence, though both sides of equation \rr{GeneralI11Expansion} can
be assumed to depend arbitrarily on $\s\uu 0$, as consistent with the
equations of motion and the Virasoro constraints.


Starting with \rr{PSL}, and motivated by \rr{GeneralI11Expansion}, we
invoke the following change of variables
\bbb
\blat = \phi + {1\over 4} \log \biggl ( {{|D|}\over{24\pi}} \cc {{\binv\lp{22}}\over{\m\uu 3 \cc \m\pr}} \cc  \cc   \biggl )\ , \qquad
\phi = \blat - {1\over 4} \log \biggl ( {{|D|}\over{24\pi}} \cc
{{\binv\lp{22} }\over{\m\uu 3 \cc \m\pr}} \cc \cc   \biggl )\ ,
\een{ScalingRegionCOVA}
along with the coordinate rescaling
\bbb
\dha{\s}\uu 1 = \binv\lp{22}\uu{{1\over 4}} \cc 
\left( {{|D|}\over{24\pi}} \right)\uu{-{1\over 4}} \cc \m\uu{{1\over 4}} \cc \m\pr{}\uu{-{1\over 4}} \cc 
\s\uu 1\ , \qquad \s\uu 1 \equiv \binv\lp{22}\uu{-{1\over 4}} \cc \left( {{|D|}\over{24\pi}} \right) \uu{{1\over 4}} \cc \m\uu{-{1\over 4}} \cc \m\pr{}\uu{{1\over 4}} \cc 
\dha{\s}\uu 1\ .
\een{ScalingRegionCOVB}
As with \rr{BulkLiouvShift}, this field redefinition and coordinate transformation make sense
only if both $\m$ and $\m\pr$ are nonzero. 
These transformations have been performed so 
that the unique time-independent classical solution 
for the shifted field $\blat$ approaches a fixed limit 
in the scaling region of fixed $\dha{\s}\ll 1$, as $|X|\to \infty$.

Of course, we can extend this to a rescaling of both worldsheet coordinates by taking
\bbb
\dha{\s}\uu 0 = \s\uu 0\ ,
\eee
so that
\bbb
\pp\ll 1 &=& \binv\lp{22}\uu{{1\over 4}} \cc \left( {{|D|}\over{24\pi}} \right)\uu{-{1\over 4}} 
	\cc \m\uu{{1\over 4}} \cc \m\pr{}\uu{-{1\over 4}} \cc \pp\ll{\dha\s^1} \ ,
\nn\\
\pp\ll 0 &=& 
\pp \ll {\dha\s^0} + \hh \cc {{\binv\lp{23}}\over{\binv\lp{22}}} \cc
\dha{\s}\uu 1 \pp\ll{\dha{\s}\uu 1} \ .
\eee
The redefinition of the $\s\uu 0$ derivative from a fixed-$\s\ll 1$ to a fixed-$\dha{\s}\ll 1$ partial derivative
does not affect the leading $J$-scalings.  For instance, we have
\bbb
\pp\ll 0 \phi = \pp\ll{\dha\s^0}\blat  - \hh {{\binv\lp{23}}\over
{\binv\lp{22}}} + \hh \cc {{\binv\lp{23}}\over{\binv\lp{22}}} \cc \dha{\s}\uu 1
\pp\ll{\dha{\s}\uu 1} \blat \ .
\een{TransformOfLiouvilleTimeDeriv}
The latter two terms are order $J\uu 0$ and are subleading relative to the $J$-scaling of the $\s\uu 0$-derivative, which we will see
is generally $O(J\uu{{1\over 4}})$ (see eqn.~\rr{FreqScaling}).

\heading{Classical solution}
In the classical ground state (i.e., the helical solution), the solution for
$\blat$ is always time-independent:
\bbb
\partial\ll{\dha{\s}\ll 1}\sqd \cc \blat = {{\dha{\s}\ll 1\uu 4}\over
4} \cc \expp{2\blat} - \expp{- 2 \blat} \qquad {\rm (ground~state)} \ .
\een{HelicalEOM}
The boundary condition for this equation states that $\blat$ obeys the
Neumann condition at $\dha{\s_1} = 0$ and continues smoothly
to all values of $\s_1 = O(1)$.  
Let us now define $\plat$ as the classical ground state solution for $\blat$.
For $\plat\pr(0) = 0$, by adjusting the initial value $\plat(0)$,
it is easy to see that the solution goes
to $+\infty$ at finite $\dha{\s}\ll 1$ for $\plat(0) > \plat\uprm{crit.} (0) $, 
and to $-\infty$ at finite $\dha{\s}\ll 1$ for $\plat(0) < \plat\uprm{crit.} (0) $,
where $\plat(0)\uprm{crit.}$ is some critical initial condition 
lying between the two singular trajectories. In other words, the only
value of $\plat$ compatible with the boundary condition 
and the existence of a smooth solution is $\plat(0) = \plat\uprm{crit.}(0)$.
It is straightforward to numerically determine 
the value of $\plat\uprm{crit.}(0)$.
For the ground-state classical solution,
\bbb
\plat(0) = 0.4067\ ,
\een{PhiBdyVal}
so we find that the full boundary value of $\Phi$ is
\bbb
\Phi(0) = 0.4067 - {1\over 4} \log \biggl ( {{|D|}\over{24\pi}} \cc {{ \binv\lp{22} }\over{\m\pr\m\uu 3}} \cc   \biggl )\ ,
\een{AsymptoticFormPhiNearBdy}
at large $J$.  
Note that the specific value \rr{PhiBdyVal} depends on the details of the perturbation of the Liouville theory.  A different form for the perturbation,
e.g., a different set of $a\ll q$ in the ansatz parametrized in \rr{GeneralX11Model}, would give the same coefficient of
$\log(\binv\lp{22})$ in equation \rr{AsymptoticFormPhiNearBdy}, but a different constant term.

%
%

\heading{Frequencies of normal modes}
Now let us estimate the frequencies of normal modes of $\tha{\phi} \equiv  \blat - \plat$, localized near
the boundary.  Such modes are of the form
\bbb
\tha{\phi} = \expp{i \o \dha{\s}\uu 0} \cc f(\dha{\s}\uu 1)\ ,
\een{NormalModeAnsatz}
where the functional dependence of $f(\dha{\s}\uu 1)$ is fixed
in the $J\to \infty$ limit.  At large $J$, the second
and third terms in \rr{TransformOfLiouvilleTimeDeriv} are negligible,
so the linearized equation of motion for the $\tha{\phi}$ fluctuation can only be satisfied if $\o$ scales as $J\uu{{1\over 4}}$:
\bbb
\o = O(J\uu{{1\over 4}})\ .
\een{FreqScaling}

Now we retain only the leading (first) term in 
\rr{TransformOfLiouvilleTimeDeriv}, and
use \rr{GeneralI11Expansion} to get a leading-order 
action for the shifted Liouville field $\blat$ in the 
scaling region $\s\ll 1 \sim \binv\lp{22}\uu{-{1/ 4}}$.  The action \rr{PSL} becomes
\bbb
&&{\cal L} =
\biggl ({{ |D| }\over{24\pi}} \biggr) (\pp\ll{\dha\s^0}\blat)\sqd 
\nn\\
&&\kern20pt - \left( {{\binv\lp{22} \cc |D| }\over{24\pi\cc }} \right)\uu{{1\over 2}} \cc \left( {\m\over{\m\pr}}\right)\uu{{1\over 2}}\cc  \biggl \{ 
(\pp\ll{\dha\s^1}\blat)\sqd + \expp{- 2 \blat} 
+  {{\dha{\s}\ll 1\uu 4}\over{4}} \cc \expp{2\blat}  \cc \biggl \} + O(|X|\uu {-1})\ .\cc\cc\cc
\nn\\
&&
\een{BdyRegAction}
So, at large $J$, the term
$\pp\ll 0$ is approximated by $\pp\ll{\dha\s^{0}}$,
and the equation of motion becomes
\bbb
\pp\ll{\dha\s^{0}}\sqd \blat = \binv\lp{22}\uu{\hh} \cc \left( {{ |D| }\over{24\pi\cc }} \right) 
   \uu{-{1\over 2}} \cc \left( {\m\over{\m\pr}} \right)\uu{{1\over 2}}\cc \cc \biggl \{ \cc \pp\ll{\dha\s^1}\sqd \blat 
+  \expp{- 2 \blat}
-   {{\dha{\s}\ll 1\uu 4}\over{4}} \cc \expp{2\blat}  \cc \biggl \}\ .
\eee
If we rewrite quantities in terms of $\plat = \blat - \tha{\phi}$,
then the normal mode equation for the mode
$f$ in \rr{NormalModeAnsatz} takes the form
\bbb
\o\sqd   f(\dha{\s\uu 1}) =  
\Theta \cdot f(\dha{\s\uu 1}) \ ,
\eee
where
\bbb
\Theta \equiv
\binv\lp{22}\uu{\hh} \cc \biggl ( {{\m}\over{\m\pr}} \biggl ) \uu{{1\over 2}} \cc \left( 
{{|D|}\over{24\pi}} \right) \uu{-{1\over 2}} \biggl \{ \cc -
 \pp\ll{\dha{\s}^1}\sqd
 + 2  \cc \expp{- 2 \plat}   +
  {{\dha{\s}\ll 1\uu 4}\over{2}} \cc \expp{2\plat}  \cc \biggl \}\ .
\een{ExplicitGaussOperator}



\heading{Quantum perturbation theory at large $J$}
Prior to the rescaling executed above, it looked as
though the theory contained any number of operators that could become
singular at the boundary.  We now see, however, that there is indeed a controlled 
perturbative expansion of the effective Lagrangian near the boundary in $\tha{\phi}$
propagators and vertices at large $J$, by using \rr{ScalingRegionCOVA} and \rr{ScalingRegionCOVB} in
the action.  Translational invariance near the boundary is
strongly broken, so the Gaussian terms for the
propagator are position-dependent, but, even so,
their $J$-scaling is simple and can be read off directly from the
Lagrangian for $\tha{\phi}$.  
That is, the Gaussian action has the form
\bbb
{\cal L}\ll{{\rm Gaussian}}\upp{\tha{\phi}} =\left( {{ |D| }\over{24\pi}} \right)  
  \left( \cc \dot{\tha{\phi}}\sqd + \tha{\phi}\, \Theta\uprm{gauss}\, \tha{\phi} \right)\ ,
\eee
where $\Theta\uprm{gauss}$ is just the operator in
\rr{ExplicitGaussOperator}.  
Therefore, the $\tha{\phi}$-propagator is always dominated by a 
position-dependent mass-squared term scaling as $\binv\lp{22}\uu{1/2}$,
which goes as $J\uu{\hh}$.  
Thus, the on-shell frequencies of near-boundary
modes will always scale as $J\uu{{1/ 4}}$.

\subsection{From UV operators to effective boundary operators}
\label{MysteriousCorrectionBoundingSubSec}
Now we would like to show that {\it any} gauge invariant boundary operator
in the microscopic theory,
or likewise any bulk operator in the scaling region $\dha{\s\ll 1} =O(1)$, goes over to an
operator satisfying the $\binv\lp{22}$-dressing rule on the boundary in
the effective theory.

To begin, it is useful to write the full expansion of the bulk operator $\inv\ll{11}$ near the boundary:
\bbb
\inv\ll{11} = \sum_{j=2}^\infty {\cal O}\upp j\ll{11} \s_1^j = \sum_{j=2}^\infty \kst\uu j\cc
{\cal O}\ll{11}\upp j \binv\lp{22}\uu{- j/4} \cc \dha{\s}_1^j\ , 
\eee
where
\bbb
 {\cal O}\ll{kl} \upp j &\equiv& {1\over{j!}} \cc \pp\ll{\s\uu 1}\uu j \inv\ll{kl} \biggl | \ll{\s\uu 1 = 0}\ ,
 \eee
 and $\kst$ is a numerical constant given by
 \bbb
 \kst &\equiv&  \left( {{|D|}\over{24\pi}} \right) \uu{{1\over 4}} \cc \m\uu{-{1\over 4}} \cc \m\pr{}\uu{{1\over 4}}\ .
\eee
The ${\cal O}\ll{kl} \upp j$ consist of operators of the form $\binv\lp{pq}$ with $p+q = j+k+l $.

Now, we consider the most general monomial perturbation of the Lagrangian
respecting Poincar\'e and worldsheet scale invariance:
\be
{\cal L}_{\rm pert} =
 e^{M\phi} \cc \prod_{p,q\geq 1} \inv\ll{pq}^{N_{pq}} \cc \prod_{r\geq 1} (\pp^r \phi)^{K_r} \ ,
\label{GenericUVOperator}
\ee
where $M$, $N_{pq}$ and $K_r$ are some arbitrary exponents.
For this contribution to be of mass dimension two, we require
\be
-M + \sum_{p,q \geq 2}(p+q) N_{pq} + \sum_{r \geq 1} r K_r = 2\ .
\ee
Using this restriction to eliminate $M$, let us rewrite ${\cal L}_{\rm pert}$ as
\be
{\cal L}_{\rm pert} = e^{-2\phi} \cc \prod_{p,q \geq 1} 
\cc {\cal S}\ll{pq}\uu{N\ll{pq}} \cc \prod_{r \geq 1} \cc {\cal T}\ll r\uu{K\ll r}\ ,
\ee
where
\be
{\cal S}\ll{pq} \equiv \exp{(p+q) \cc \phi} \cc \inv\ll{pq} \ , \qquad {\cal T}\ll r \equiv \expp{r \phi} \cc \pp\uu r \phi\ .
\ee
The objects ${\cal S}\ll{pq}$ and ${\cal T}\ll r$ have scaling dimension zero, so ${\cal L}_{\rm pert}$ 
is now strictly of mass dimension two.

We can now compute constraints on the $|X|$-scaling of terms in ${\cal L}_{\rm pert}$ 
in the boundary region.  In terms of the shifted field $\blat$ \rr{ScalingRegionCOVA},
\be
\exp{(p+q) \cc \phi}  \sim \binv\lp{22}\uu{-(p+q)/4} \cc \exp{(p+q) \cc \blat}  \ , 
\ee
so the ${\cal S}\ll{pq}$ contribution 
to the $|X|$-scaling of ${\cal L}_{\rm pert}$ at the boundary is $2-(p+q)/2$.
Now, the only ${\cal S}\ll{pq}$ that can potentially contribute positive $|X|$-scaling
overall are those for which $p+q \in \{2,3\}$.  $\inv\ll{11}$ scales at the boundary 
as $-\hh\cc \binv\lp{22}\cc \s\ll 1\sqd = -\hh\cc \kst\sqd  \binv\lp{22}\uu{1/2} \cc
\dha{\s}\ll 1 \sqd $, so  
\be
{\cal S}\ll{11} = \expp{2\phi} \inv\ll{11} \sim  
-{{\m\m\pr}\over 2} \cc \binv\lp{22}\uu 0
\cc \expp{2\blat} \cc \dha{\s}\ll 1\sqd 
= O(|X|\uu 0)\ .
\ee
Similarly, $\inv\ll{12}$ and $\inv\ll{21}$ at the boundary behave like 
$\pm \hh \cc \binv\lp{22} \cc \s\ll 1 \sim \pm\hh \cc \binv\lp{22}\uu{{3/ 4}} \cc \dha{\s}\ll 1$, so
${\cal S}\ll{12}$ and ${\cal S}\ll{21}$ go as  
$\pm \hh \cc \binv\lp{22}\uu 0 \cc \expp{3\blat} \cc \dha{\s}\ll 1 = O(|X|\uu 0)$.

The $\binv\lp{22}$ scaling of factors of the form ${\cal T}\ll r$ can be analyzed in a similar fashion.
The $\binv\lp{22}$-scaling of contributions from $\expp{r\phi}$ are $\binv\lp{22}^{-r/4}$,
while the dominant $\binv\lp{22}$ contributions from the $\phi$ derivatives in ${\cal T}\ll r$ descend
from $\pp_1$ derivatives, and these become $\pp_1^r = \kst\uu{-r}\cc \binv\lp{22}^{r/4}\pp_{\dha\s^1}^r$ in 
our rescaled coordinates.  So the ${\cal T}_r$ objects themselves enter 
with $\binv\lp{22}$-scaling $\binv\lp{22}^0$.  Thus, we have seen that the $\binv\lp{22}$-dressing rule is satisfied in the effective theory, 
for any operator insertion in the UV theory
of the general form \rr{GenericUVOperator}.

As a specific example, we now demonstrate that 
the boundary Liouville term in the microscopic theory descends to a quark mass 
term in the effective string theory.
In this regime, we can calculate the numerical value of the coefficient of the quark mass operator in the effective string
theory derived from the Polchinski-Strominger deformed Liouville theory, with space-filling branes in $D+1$ dimensions, and a boundary Liouville
term:
\bbb
{\cal L}\ll{\rm boundary} = \m\ll B \cc \expp{-\Phi} \to {{0.6558 \cc \m\ll B}\over{(\m\uu 3 \cc \m\pr)\uu{{1\over 4}}}} \cc 
\left( {{ |D| }\over{24\pi }} \right) \uu{{1\over 4}} \cc \binv\lp{22}\uu{{1\over 4}}\ .
\een{EffectiveQuarkMassTermCoeff}  
When we expand the $X$ field in vev plus fluctuations, the terms with fluctuations have lower $J$-scaling than the $\binv\lp{22}$ term evaluated
in the classical solution.  Therefore, the coefficient of the $J\uu{{1/ 4}}$ term in the open string mass-squared is set directly
by the coefficient in equation \rr{EffectiveQuarkMassTermCoeff}, regardless of the details of the state, so long as its excitation number
above the ground state is not parametrically large in $J$.


\subsection{Quantum corrections leave the boundary dressing rule unmodified}
\label{BdyQuantCorr}
To show that quantum corrections do not affect the dressing rule at
the nonperturbative level, all we need to establish is that the energies of the modes we are integrating out go as 
$\o\ll{\s\uu 0} \sim \binv\lp{22}\uu{{1/ 4}}$.  In a Wilsonian action, 
the dimensional suppression of a nonrenormalizable effective term is generated by 
inverse powers of the frequencies of the modes that generated the term when they were integrated out.  Bulk terms are generated by
integrating out bulk modes of the Liouville field, whose bulk mass is 
proportional to $\sqrt{|\inv\ll{11}|}$.  On the other hand, near the boundary, the dimensional
suppression of boundary operators comes from negative powers of the frequency of near-boundary modes of the
fluctuations of $\phi$.  From equation \rr{BdyRegAction} we can see that the frequencies of these modes 
are of order
  \bbb
  \o
   = O\biggl (  \binv\lp{22}\uu{{1\over 4}} \cc \left( {{|D|}\over{24\pi}} \right) \uu{-{1\over 4}} 
  \biggl )\ .
  \eee  
It follows that $\binv\lp{22}$ is indeed the operator dressing for all boundary
operators.  This is a nonperturbative statement, and the 
powers of $\o$ that appear -- and therefore the powers of 
$\binv\lp{22}$ that appear -- depend on the full nonperturbative 
dynamics of the strongly coupled conformal field theory through 
the anomalous dimensions of the operators they are dressing.
Note, however, that the form of the dressing, as opposed to its exponents, 
is the same as it is in large-$|D|$ perturbation theory: Both the classical solution for $\phi$ and
the propagator for its fluctuations, contain only powers and
logarithms of $\binv\lp{22}$.

\def\sreg {displaced-boundary regulator~}
\def\sregp{displaced-boundary regulator}
\section{A \sreg}
\label{FiniteSepReg}
To this point, we have been working from the starting
point of a 
UV-complete worldsheet theory, taking the form 
of a Liouville model coupled to the goldstone bosons $X$.  As we have seen,
this ultraviolet completion provides a natural way to regulate the effective string theory, giving the 
$\binv\lp{22}$-dressing rule for effective Neumann boundary operators,
as well as the $\inv\ll{11}$-dressing rule for bulk effective operators.
Deriving the dressing rule from a UV-complete theory, however, should
not overshadow one very important point:  The structure of these operator 
dressing rules is an {\it intrinsic} property of the effective theory. 
That is, the operator dressing rules we have demonstrated above
do not arise as an artifact of a particular UV-completion of
effective string theory.  To illustrate this point in greater detail,
we can adopt a different regulation procedure
that is not related to any
particular physical completion of the effective theory.  For instance, we can 
work instead with an artificial cutoff, which renders calculations
tractable while preserving the underlying symmetries of the worldsheet
theory.  The $\binv\lp{22}$-dressing rule again emerges ineluctably in such
schemes, so long as one takes care to preserve worldsheet 
(diff)$\times$(Weyl) gauge symmetries, as well
as the global $D$-dimensional Poincar\'e symmetry.

One such regulator is defined by replacing the boundary
of the worldsheet with a displaced timelike boundary, moved 
slightly into the interior of the worldsheet by a fixed distance $\e$.  
We refer to this scheme as the ``\sregp."
For bulk operators, the significance
of the displaced boundary is that we integrate only up to
the displaced boundary rather than the real one, excising
interaction terms from the strip at the boundary.  For boundary
operators, we integrate along the displaced boundary
rather than the real one.  For instance, for a quark
mass, we simply let the worldline of the quark run along the
displaced boundary.

The spacelike displaement $\e$ is measured with respect
to the induced metric.  This regulator is fully gauge 
invariant by construction, referring only to gauge invariant 
quantities, and approaches the bare quark action when $\e$ is taken to
zero.  

To avoid complications in the near-boundary expansion, we can define
the induced proper distance with respect to the free, rather than interacting, $X$-coordinates, 
which is indeed equivalent to turning off all bulk interactions inside the excised strip. 
This is a perfectly well-defined and gauge-invariant procedure, modulo the subtle complication of turning off the anomaly
term in the near-boundary strip.  However, the only effect of the excision of
the anomaly term is a variation of the quantum effective action by the Wess-Zumino functional integrated
over a strip of coordinate width proportional to $\e\uu{\hh}$.  This goes to zero as $\e\to 0$, and gauge invariance
is restored as the regulator is removed.  Concretely, we will see below that the effect of the near-boundary
excision of the anomaly term is equivalent to adding a gauge-invariant boundary counterterm, plus other boundary terms that vanish in the limit $\e\to 0$.

Alternatively, we can define the induced proper distance with respect to the interacting embedding coordinates.  If we do so, we must separately
regulate the bulk interactions (including the anomaly term) and remove the associated divergences with boundary counterterms again.\footnote{We give
an example of a regulator of this type in Appendix \ref{derivationOfReggeIntercept}, where we perform a fully gauge-invariant calculation 
of the asymptotic intercept.  In this calculation, the excision of the strip is not needed, as the only divergence comes from the bulk anomaly
term, which we regulate explicitly.}

In what follows, we work through three examples of interest, 
and the \sreg proposed here handles each differently.  
First, we consider boundary terms.  The
naive quark-mass operator, regulated according to this scheme, 
requires a multiplicative renormalization in order to have nonzero matrix elements in low-energy states in the limit where the
regulator is removed.  It is important
to note that this is always true when we are expanding 
in the limit where the quark mass is held fixed, at any
finite value, and $J$ is taken to infinity.  
Other boundary operators, such as the integrated geodesic curvature, or
the proper acceleration operator, are proportional to the 
quark-mass operator at leading order in $J$.

Second, we investigate fully gauge invariant bulk terms. 
For one such example, that of the induced-curvature-squared 
term, we compute the leading divergence as the size
of the \sreg is taken to zero.  Here again, the divergence 
is proportional to the quark mass term, with a coefficient 
scaling as $\e\uu{-{5/ 2}}$.  

Finally, we consider the anomaly term itself. The naive
near-boundary regulation of this 
term does not result in a gauge-invariant theory, because
the anomalous transformation of the free theory is cancelled 
only by the integral of the PS term over the full worldsheet.  
In Section \ref{StripExcisAnom}, we show that gauge invariance
is restored in the limit $\e\to 0$, by decomposing the 
integral of the PS term in the strip as a gauge invariant operator
(proportional to $\binv\lp{22}\uu{{1/ 4}}$) and terms
that vanish as $\e\to 0$.

More generally, and returning to the central lesson of the dressing
rule, there is only one invariant perturbation of the action of order $J\uu{{1/ 4}}$, 
and all gauge-invariant operators are just proportional to a
single, linearly independent operator, $\binv\lp{22}\uu{{1/ 4}}$, at order $J\uu{{1/ 4}}$.  This operator
should be thought of as just the identity, dressed with a line element along the boundary, expanded near
the boundary and renormalized multiplicatively to give a finite and nonzero value.


\subsection{Definition of the \sreg scheme}

To begin, we attach the worldline of the quark to
the string worldsheet, separated from the boundary by a fixed spacelike induced proper distance $\e$.
That is,  the distance from the quark to the boundary is computed with respect to the induced metric on the 
worldsheet,
\bbb
G\uprm{ind}\ll{\s\uu a \s\uu b} \equiv \pp\ll a X\cdot \pp\ll b X\ .
\eee
The simplest gauge-invariant characterization of a
trajectory near the boundary is defined to be the set of
interior points separated from the boundary by extremal spacelike geodesics of length
$\e$ with respect to the induced metric.\footnote{In particular,
the separation is characterized by spacelike geodesics of maximal length (in Lorentzian signature),
extending from an interior reference point to the boundary.  See Appendix \ref{GeodDistBdyFunc} for
further details on the maximal geodesic in the near-boundary region.}  
The spacelike geodesics of interest
connecting interior points to the boundary are simply slices of
constant $\s\uu 0$, 
parametrized by $\s\uu 1$.  That is, the set of points lying at a fixed spacelike geodesic
distance $\e$ from the boundary is just a trajectory of constant $\s\ll 1$:
\bbb
\s\ll 1 = \tilde{\s} = (\s\uu 0-{\rm independent})\ .
\een{SepTrajHelical}
With this, we can work out the actual value of the coordinate location
$\tilde{\s}$ of the near-boundary trajectory, in terms of the induced proper distance $\e$.

Concretely, the induced proper distance is
\be
\e = \int_0^{\tilde\s} d\s \sqrt{G\uprm{ind}_{\s\s}} = \int_0^{\tilde\s} d\s \sqrt{\pp_\s X^\m \pp_\s X\ll\m}
 =  \int_0^{\tilde\s} d\s\sqrt{-2 \inv\ll{11}}\ .
\label{epsDef}
\ee
Recalling the boundary expansion of $\inv\ll{11}$ in
eqn.~\rr{GeneralI11Expansion}, we obtain
\be
\e = \int_0^{\tilde\s} d\s \sqrt{\binv\lp{22}\s^2 + O(\s^3)} =
\frac{1}{2} \sqrt{\binv\lp{22}} \cc \tilde\s^2 + O(\tilde\s^3) \ ,
\label{epsSol}
\ee
or
\be
\tilde\s = \sqrt{2\e} \cc \cc  \binv\lp{22}^{-\frac{1}{4}} + O(\e)\ .
\label{SigmaTildeSol}
\ee

The expressions in this section are again
operator identities, and not simply properties of
the classical helical solution; this point warrants some discussion.
The expansion of $\inv\ll{11}$ near the boundary in \rr{epsSol} receives corrections due to bulk interactions, 
just as it did in the Polchinski-Strominger-Liouville UV completion,
as discussed above \rr{GeneralI11Expansion}.
In the UV-complete theory, the bulk interactions were nonsingular near the 
boundary, and their near-boundary expansion was unproblematic.  
Strictly within the effective theory, however, we have bulk operators that are explicitly singular near the 
boundary, where these singularities are excised by the explicit cutoff at fixed 
induced proper length $\e$ from the boundary.

For instance, there are terms coming from the PS correction and
from other corrections to the free-field action and its constraints.  
The d'Alembertian on $X\uu\m$ contains singular terms such as
$\b\apr {{\inv\ll{21} \inv\ll{12}\sqd}\over
{\inv\ll{11}\uu 4}}\cc\pp\ll + X\uu\m$, leading to corrections to the RHS of \rr{GeneralI11Expansion} of the 
form $\b\apr{{\inv\ll{21}\sqd \inv\ll{12}\sqd}\over{\inv\ll{11}\uu 4}}\s\ll 1\sqd$, for example, which, near the boundary, 
behave as 
\be
\b\apr{{\inv\ll{21}\sqd \inv\ll{12}\sqd}\over{\inv\ll{11}\uu 4}}\s\ll 1\sqd \biggr|_{{\rm boundary}} \to
{{\b\apr }\over{\s\ll 1\sqd}}\ .
\label{NearBdyExpOfPSTermTakeOne}\ee
As noted above, we can deal with such singularities either by introducing a separate UV cutoff for the bulk
interactions, or by simply defining the induced geodesic distance with respect to an embedding in which
the embedding coordinates satisfy the free, rather than interacting, EOM 
inside the excised strip.  We take the latter approach 
in this section.

It is also important to note that, while $\e$ is a number, $\tilde{\s}(\s\uu 0)$ is actually
an operator, denoting the coordinate position at which
the induced proper distance from the boundary is equal to $\e$.
The $\e$-dependent terms coming from the regulator 
enter in a series with hierarchical $J$-suppression, determined by how
many powers of the dressing operator $\binv\lp{22}$ they carry in the denominator. 

There is a second kind of large-$J$ suppression 
associated with the expansion of any given operator into a classical background and quantum fluctuations, 
i.e.,~
\bbb
X\uu\m\equiv Y\uu\m + E\uu\m\ll{{\rm helical}}\ ,
\een{YFlucDef}
where $E\uu\m\ll{{\rm helical}}$ is the helical classical solution, whose invariants (such
as $\inv\ll{11}$) are time-independent.   For any
given configuration in the path integral, the
length of the extremal geodesic integrated out to
$\s\ll 1 = \tilde{\s}$ in the boundary
scaling region contains corrections such
as $\hh\sqrt{B\lp{22}} \tilde{\s}\uu p {\cal O}\ll p$,
where ${\cal O}\ll p$ is an operator of dimension
$p$ made of derivatives of $Y$, e.g.,~${\cal O}\ll p \ni (\pp Y)\uu p$.  Since
$\tilde{\s}$ depends on the configuration and $\e$ does
not, it is better to express the corrections to \rr{SigmaTildeSol} in 
terms of fixed-$\e$ quantities,
\bbb
\tilde{\s} =  \sqrt{2\e} \cc \cc  \binv\lp{22}^{-\frac{1}{4}} \cc\left( 1 + O(\e\uu{p/2}
B\lp{22}\uu{-p/4} {\cal O}\ll p) \cc\right) \ .
\een{SigmaTildeSol2}
Note, in particular, that the correction terms in the equation above will be suppressed both by 
$J$ and $\e$.


\subsection{Boundary operators}
We now expand some simple gauge-invariant operators near the boundary,
with the boundary proximity defined in a gauge-invariant way,
based on the induced proper distance.  The key result here is that gauge invariant local operators have near-boundary expansions
that go as $({\rm const.}) \cc J\uu 0$, with the constant depending
only on the regulator and the Hamiltonian of the system, and
not at all on the state.  In other words, the leading coefficient can be theory-dependent, but not state-dependent within a given effective theory.
After being multiplied by the induced boundary line element to make a gauge-invariant perturbation to the action, gauge-invariant
terms scale as $({\rm const.}) \cc \binv\lp{22}\uu{{1/ 4}}$, where the constant factor is independent of the state.

\heading{Quark mass operator from naive quark action}
\label{NaiveQMop}
The simplest case to consider is the operator $\binv\lp{22}$, in which we
simply take the identity, multiply it by a regulated induced line element
with a coefficient we can think of as a bare quark mass term.
The naive quark mass term thus appears as
\be
S\lprm{quark~mass} = M\uprm{bare} \cc \int \cc ds\uprm{induced} = M\cc \int \cc d\r \cc \sqrt{ - {{dX\uu\m}\over{d\r}} \cc {{dX\ll\m}\over{d\r}} }\ ,
\label{QMTermNaive}
\ee
where the integral is now understood to be taken over a timelike 
trajectory near the boundary, and where the separation from the boundary 
is parametrized by the cutoff $\e$.  
Taking the integral over an exactly lightlike boundary leads to a
singular worldsheet Hamiltonian; the nonsingular operator is obtained 
by rescaling the naive term.  Stated another way, after fixing a gauge-invariant 
scheme parametrized by $\e$ to regularize the term, we make an $\e$-dependent readjustment of
the bare parameter $M$:
\bbb
S\lprm{quark~mass} = M\uprm{bare} (\e) \cc  \int \cc ds\uprm{induced} \ .
\eee

As noted, the renormalization can depend on the scheme, and on the
parameters of the worldsheet Hamiltonian, but cannot depend on the 
state of the system.  Therefore, it is simplest to determine the
scaling by simply inserting the quark mass term into the helical
solution.  
Evaluated on this trajectory, the line element in \rr{QMTermNaive} goes as
\bbb
 d\r \cc \sqrt{ - {{dX\uu\m}\over{d\r}} \cc {{dX\ll\m}\over{d\r}} } &=& d\t \cc \sqrt{- \dot{X}\sqd}  = d\t \cc \sqrt{- 4 \inv\ll{11}} 
 \nn\\
& \simeq &d\t \cc \sqrt{2 \binv\lp{22} \cc \tilde{\s}\sqd} =2\cc d\t\cc \e\uu{\hh} \cc \binv\lp{22}\uu{{1\over 4}}\ ,
\een{ThisEqRef}
where we have used the Virasoro constraints 
to substitute $-\dot{X}\sqd \sim - 4 \cc \inv\ll{11}$, 
and equation \rr{GeneralI11Expansion} to approximate 
$\inv\ll{11}$ by $-\hh \cc \binv\lp{22} \cc \tilde{\s}\sqd$.  Thus, 
the bare mass $M\uprm{bare}(\e)$ must be scaled as $\e\uu{-\hh}$ 
to recover a finite operator when $\e$ is taken to $0$:
\bbb
 M\uprm{bare} (\e) \sim c\ll{\rm quark} \cc \e\uu{-\hh}\ .
 \label{mbare}
\eee
We have discarded subleading terms on
the RHS of \rr{ThisEqRef} such as $\e\cc {{\binv\lp{23}}/{\binv\lp{22}\uu{{3/ 2}}}}$. 
In general, boundary operators contributing
to $\tilde{\s}$ have scaling dimension $-1$ and so must
come dressed with an $(\e\sqd / \binv\lp{22})\uu{{{p + q - 4}\over 4}}$ for every $\binv\lp{pq}$ in the numerator.  The expansion
of the formula for $\tilde{\s}$ in boundary operators 
is therefore an expansion in (fractional) positive powers of ${{\e\sqd}/{J\apr}}$.

Note that all throughout, we have taken the physical effective
quark mass term, defined as the $J\uu{{1/ 4}}$ contribution
to the mass-squared, to be zero prior to the perturbation \rr{QMTermNaive}.  
In the presence of a nonzero quark mass coefficient, 
the boundary of the worldsheet is slightly timelike at finite $J$, and the operator renormalization \rr{mbare} 
may be deformed.\footnote{We thank S. Dubovsky and V. Gorbenko for 
discussions of this point.}  However, the basic organization 
of boundary operators
itself does not depend on the value of $c\ll{\rm quark}$: The basis of boundary operators is still the set of 
arbitrary polynomials of $X$ and its derivatives, dressed with powers 
of $\binv\lp{22}$, so long as $c\ll{\rm quark}$ is taken fixed
and independent of $J$ as $J$ is taken to infinity.\footnote{Note that it is also possible to take a scaling limit where the coefficient of
the $\binv\lp{22}\uu{{1\over 4}}$ term is scaled as $J\uu{{3\over 4}}$, 
in which the velocity of the endpoint stays finite and the
organization of boundary operators changes. See, for
instance, the discussion of rotating strings at large $J$ and 
fixed endpoint velocity in \cite{Sonnenschein:2014jwa}.}

\heading{Boundary operator: Geodesic curvature}
Another example of a boundary operator is the geodesic curvature.
Taking $\g$ to be a curve of fixed $\s$, with unit tangent vector $u = d\g/ds$, we have
\be
u &=& 
\frac{d\g}{ds} = \frac{(\t'(\rho),\s'(\rho))}{|g_{ab} \s^{a\prime}(\rho) \s^{b\prime}(\rho)|^{1/2}}
\nn\\
& = & \frac{(1,0)}{\sqrt{2|\inv\ll{11}|} } \ ,
\ee
where we have chosen the specific parameterization $\t=\rho$.  The geodesic curvature squared is 
\be
\kappa^2 = u^c u^d D_c u^a D_d u^b  g_{ab} \ .
\ee
Expressing this quantity in terms of the embedding coordinates via the induced
metric, we obtain
\be
\kappa^2 = \frac{1}{4}(u^0)^4 g^{11} (g_{00,1})^2  = \frac{(\dot X \cdot {\dot X}')^2}{(\dot X \cdot \dot X)^3} \ .
\ee
Taken to the boundary cutoff $\epsilon $,
\be
\k\sqd = \frac{1}{\binv\lp{22} \s^4} \biggr|_{\epsilon} \ ,
\ee
where $\tilde \s = \sqrt{2\epsilon} \binv\lp{22}^{-1/4} + O(\e)$, we obtain
\be
\kappa^2 \longrightarrow \frac{1}{4\epsilon^2} \ .
\ee
The matrix elements of this operator are theory-dependent, regulator-dependent, and UV singular for massless endpoints, but
have the same value at order $J\uu 0$ for every state in the low-energy Hilbert space.  Treated as integrated perturbations,
they scale as $J\uu{{1/ 4}}$ after multiplication by the 
induced boundary line element $\binv\lp{22}\uu{{1/ 4}}$.

\heading{Boundary operator: Proper acceleration}
Let us now consider the proper acceleration,
\be
a^a = X^a_{,\rho\rho} + \Gamma^a_{bc} X^b_{,\rho} X^c_{,\rho}\ ,
\ee
where it can be shown by similar methods that, near the boundary,
\be
a^a_{\rho\rho} a^a_{\rho\rho} (g^{\rho\rho})^2 
      = \frac{(\ddot X)^2}{(\dot X^2)^2} 
      = \frac{\binv\lp{22}}{(\binv\lp{22} \tilde \s^2)^2} 
      = \frac{1}{4\epsilon^2} \ .
\ee

We have now expanded several gauge-invariant operators in a near-boundary 
expansion, where the distance to the boundary is defined in a
gauge-invariant way.  
With the regulation parameter $\e$ held fixed, we have found that
gauge-invariant scalar operators scale as $J\uu 0$ in the large-$J$
limit, with a coefficient that depends only on the parameters of the
theory and on the regulator, and not on the individual state of the
system.  In particular, the coefficient can be read off directly 
from its value in the helical state.  After multiplication by the 
regulated line element, which scales as $\binv\lp{22}\uu{{1/ 4}}$, 
these operators contribute to the action at leading order 
as $J\uu{1/ 4}$, with a state-independent coefficient.  
The operators in the examples above demonstrate the dressing rule concretely, 
in the context of a particular short-distance cutoff.

\subsection{Bulk operators}
As an example of a bulk operator analysis in the \sreg
scheme, consider the curvature-squared of the induced metric:
\be
\tilde R^2 d\mu_{\rm Induced} = 8 \frac{({\hat \inv\ll{22}})^2}{|\inv\ll{11}|^3} d\s^0 d\s^1 \ ,
\ee
where, as
in \cite{Hellerman:2014cba}, we define $\widehat{\inv}\ll{22}$ to be the
Weyl-covariant version of $\inv\ll{22}$: 
\bbb
\widehat{\inv}\ll{22} \equiv \inv\ll{22} - {{\inv\ll{12} \inv\ll{21}}\over{\inv\ll{11}}}\ .
\een{I22hatDef}
Near the boundary, this becomes
\be
\frac{(\widehat \inv\ll{22})^2}{\inv\ll{11}^3} \biggr|_{\rm bdy} =
-\frac{1}{2}\binv\lp{22}^{-1} \s^{-6} \ ,
\ee
so that
\be
\int_{\tilde\s}^{\s_{\rm far}} \tilde R^2 d\s &=& \int_{\tilde\s}^{\s_{\rm far}} 8 \frac{(\widehat \inv\ll{22})^2}{\inv\ll{11}^3} d\s 
\nn\\
&=&  \frac{4}{5} \binv\lp{22}^{-1}\left(\s_{\rm far}^{-5} - \tilde\s^{-5}\right)
\nn\\
&=& \frac{4}{5} \binv\lp{22}^{-1}\left(\s_{\rm far}^{-5} - \frac{1}{(2\e)^{5/2} \binv\lp{22}^{-5/2}} \right)
\nn\\
&=& -\frac{4}{5} (2\e)^{-\frac{5}{2}} \binv\lp{22}^{\frac{1}{4}}\ .
\label{FinalAns}
\ee

Note that we have dropped subleading divergent terms in the near-boundary expansion, which
give rise to operators of lower $X$-scaling and less divergent $\e$-scaling than the leading term \rr{FinalAns}.
For instance,  including
the $\s\uu{-4}$ term in the near-boundary expansion of
$\inv\ll{11}\uu{-3}$ leads to a divergence 
proportional to ${{\binv\lp{33}}/{\binv\lp{22}\uu{5/ 4}}} \cc \e\uu{-
{3/ 2}}$ in the integrated curvature-squared term, which scales as $J\uu{-{1/ 4}}$.  In addition, there are similar corrections
coming from the subleading terms in the expansion of $\tilde{\s}$ in powers of ${{\e\sqd}/{J\apr}}$, discussed below \rr{SigmaTildeSol}.

In particular, any $\e$-independent terms in the integral must scale as ${1/({J\apr})}$ at most, simply by dimensional analysis.  Thus,
while the UV-divergent counterterms may have $J$-scaling as large as $J\uu{{1/ 4}}$, the observable, finite contribution of 
the curvature-squared operator to the mass-squared of the open string state is no greater than $O(J\uu{-1})$.  More generally,
a higher-derivative bulk operator with $|X|$-scaling $|X|\uu{-p}$ 
may generate UV divergences going as $\e\uu{p\pr - p}{\cal O}\lp {p\pr}$, where
${\cal O}\lp {p\pr}$ is an operator with $|X|$-scaling $-p\pr$.  Finite terms can only scale as $O(J\uu{-{p/ 2}})$ at most.  
This is the basis for the perturbativity of the properly renormalized effective worldsheet theory in the $1/J$ expansion.

\subsection{Anomaly term}\label{StripExcisAnom}
The \sreg treatment of the anomaly term must be handled
somewhat more cautiously than the corresponding regulation of 
gauge-invariant bulk terms.  The naive regulation of the anomaly 
term does not result in a gauge-invariant path integral, because
the anomalous transformation of the free theory is cancelled 
only by the integral of the PS term over the full worldsheet.  

One way to proceed is to use a 
modified PS term that is finite at the boundary while
preserving exact gauge invariance.  Such a term can 
be constructed easily in the general framework of \cite{Hellerman:2014cba}. 
This approach yields a notably different situation, with two 
classes of bulk terms.  One class admits gauge-invariant 
terms of definite $X$-scaling, which are integrated over the
worldsheet with the boundary strip removed.  
The other class of bulk terms have inhomogeneous $X$-scaling by necessity.
They behave like a Liouville term under Weyl transformations, and are
integrated over the entire strip.  

Another approach is simply to integrate the 
standard, unmodified anomaly term over the worldsheet with 
the excised strip, and notice that the only term with a 
nonpositive $\e$-scaling is the quark mass term, which is 
gauge invariant.  Modulo terms that vanish as $\e\to 0$,
the strip-excision prescription for the anomaly term 
results here in a gauge-invariant and finite quantum effective 
action (because one can renormalize the sole divergence 
with a gauge-invariant counterterm).


Near the boundary, it can be shown that 
\be
{\cal L}_{\rm PS} = -\frac{\b}{2\pi} \frac{1}{\s^2} + O(\s^0) \ .
\ee
The \sreg integral then yields
\be
\int_{\tilde \s}^{\s_{\rm far}} {\cal L}_{\rm PS}\, d\s =
-\frac{\b}{2\pi} \frac{\binv\lp{22}^{1/4}}{\sqrt{2\e}} + O(\e^0) \ .
\ee
This defines the counterterm to be added to the action.  Although this regulator does
not, strictly speaking, preserve gauge invariance for $\e \neq 0$, we
proceed by adding
\be
\Delta {\cal L} = \left( c_{\rm divergent} + c_{\rm finite}\right)
\binv\lp{22}^{1/4} \ ,
\ee
with 
\be
c_{\rm divergent} = \frac{\b}{2\pi} \frac{1}{\sqrt{2\e}} \ .
\ee
We are free to choose $c_{\rm finite}$, as long as it is $\e$-independent.
The result is a gauge invariant, finite path integral in the limit
$\e \to 0$.  This concretely illustrates the restoration of gauge invariance in the limit $\e\to 0$: The near-boundary region's contribution
to the anomaly action is seen explicitly to be equal to a (gauge-invariant) boundary term, plus contributions scaling as positive powers of $\e$.


\section{Conclusions and physical consequences}
\label{conclusions}
We have proposed an operator dressing rule for Neumann boundaries in effective string theory in which 
the boundary operator ${\binv}\lp{22}$ plays the role of the unique monomial in $X$ occurring to negative
or fractional powers, analogous to $\inv\ll{11}$ for bulk
operators.  Having motivated the rule via a number of heuristic
arguments, we then explicitly derived the ${\binv}\lp{22}$ dressing rule
for Neumann boundaries in detail in a particular UV completion of
the effective theory.
To avoid any reliance on the specific UV completion, we also demonstrated
the persistence of the dressing rule using an artificial regulator directly in the effective theory.

There is at least one important physical consequence of this analysis.
Having established the dressing rule for Neumann boundaries, we can
explain in detail the universality of the asymptotic Regge intercept
for mesons in planar QCD with massless, bosonic quarks, first calculated in \cite{PRL}.
(For completeness, we include an explicit, gauge-invariant calculation of the asymptotic Regge 
intercept in this theory -- for bosonic quarks on the leading trajectory, in the planar approximation --
in Appendix \ref{derivationOfReggeIntercept}.)
The argument stems from the fact that
all bilinear invariants of the embedding coordinates $X$ at the
boundary are of the form 
\be
\binv\lp{pq} \equiv \pp\ll 0\uu p X \cdot \pp\ll 0\uu q X\ ,
\ee
and, by the dressing rule, boundary operators are spanned by the set
\be
\prod\ll{i} \cc \binv\lp{p\ll i q\ll i } / \binv\lp{22}\uu k\ .
\ee
Now, consider only boundary operators of marginal scaling dimension.
If an ``undressed'' operator (the numerator) has 
dimension 
\be
\D \equiv \sum\ll{i} p\ll i + q\ll i\ ,
\ee 
then the dressing, under the requirement of conformality, is 
\be
\binv\lp{22}\uu{ - (\D - 1) / 4}\ .
\ee
Thus, to have positive or zero $X$-scaling, the
undressed operator must have \bdol{\D \leq 5}.  
The operators $\binv\lp{11}$ and $\binv\lp{12}$, for instance, vanish as 
independent operators because they are proportional
to free-field stress tensors and first derivatives 
thereof.  Meanwhile, the only marginal operator with 
$\Delta = 5$ is $\binv\lp{23} / \binv\lp{22}$,
which is a total derivative along the boundary.  Thus, after 
modding out by Virasoro descendants, the only marginal 
operator with nonnegative $X$-scaling is the quark mass
operator, corresponding to $\Delta = 4$.
There are no operators scaling as $J\uu 0$, 
so the $J\uu 0$ term in the expansion of the quantum effective action
is indeed universal.  In particular, the order $J\uu 0$ term in the expansion 
of the mass-squared of the meson is independent of the
details of the theory, beyond the basic assumptions of $D$-dimensional 
Poincar\'e invariance and the restriction that the Nambu-Goldstone
bosons constitute the only infinite-range excitations
on the string worldvolume.

It is also worth mentioning 
what we expect to hold as corresponding dressing rules when Dirichlet boundaries are included.
In the case of strictly Dirichlet boundary conditions, the dressing rule should be formulated purely in powers of
$(\dot X')^2$.  When both Neumann and Dirichlet directions are present, we expect the appropriate dressing rule to 
be formulated in terms of $(\dot X\ll{\textsc{Neumann}})^2$, or, equivalently by virtue of Virasoro constraints, in terms of $(X'\ll{\textsc{Dirichlet}})^2$.


Looking ahead, the renormalization analysis in this paper can and should be extended to the fold singularities of rotating strings
with angular momentum in a single plane 
(see \cite{Hellerman:2014cba,PRL,Ganor:1994rm,Caron-Huot:2016icg,Sonnenschein:2015zaa} and references therein
for further discussion on this topic).
It would also be interesting to understand the origin of the Neumann 
dressing rule in the context of Natsuume's warped UV completion \cite{German:1997hc}, 
as this ties in most directly with modern holographic ideas in this arena.  


\appendix{Calculation of the asymptotic Regge intercept}
\label{derivationOfReggeIntercept}
\numberwithin{equation}{section}

In previous work \cite{PRL} we presented an abridged computation of the 
first sub-leading correction near large $J$ to the ground state energy of spinning strings, arising from contributions 
from the Casimir energy and from the Polchinski-Strominger anomaly
term discussed above.  
(The origin of the PS term in the general setting of a perturbed Liouville theory embedded 
in the Polyakov framework was further explained in
\cite{Hellerman:2014cba}.)
In this section we provide an explicit and completely gauge-invariant
derivation of these universal sub-leading corrections to the energy
spectrum for open strings.

From \cite{PRL,Hellerman:2014cba}, the ground state helical solution discussed above 
can be explicitly written as
\be
X\uu 0  & = &  2 \apr P^0 \s^0 
\nn
\\
\Zb\ll 1 &=&   i \sqrt{\frac{\apr}{2}}  \a_1^{\bar Z_1}  \biggl ( \cc
 e^{-i  \s\uu + }  + 
 e^{-i  \s\uu - }
    \biggl )\ 
\nn
\\
    \Zb\ll 2 &=&   i \sqrt{\apr\over 2} \frac{\a_2^{\bar Z_2}}{2}  \biggl ( \cc
 e^{-2 i  \s\uu + }  + 
 e^{-2 i  \s\uu - }
    \biggl )\ 
\nn
\\
Z\ll 1 &=&   -i \sqrt{\apr\over 2}  \a_{-1}^{Z_1}  \biggl ( \cc
 e^{i  \s\uu + }  + 
 e^{i  \s\uu - }
    \biggl )\ 
\nn
\\
    Z\ll 2 &=&   -i \sqrt{{{\apr}\over 2}} \frac{\a_{-2}^{Z_2}}{2}  \biggl ( \cc
 e^{2 i  \s\uu + }  + 
 e^{2 i  \s\uu - }
    \biggl )\ ,
\een{ClassicalHelical}
with
\be
\a_1^{\bar Z_1} = \sqrt{2J\ll 1}   &\quad & \a_{-1}^{Z_1} = \sqrt{2 J\ll 1} 
\nn\\
\a_2^{\bar Z_2} = 2 \sqrt{J\ll 2}  &\quad & \a_{-2}^{Z_2} = 2 \sqrt{J\ll 2} \ .
\ee
The usual classical constraint takes the form,
\be
T_{++} = -(\pp_+ X^0)^2 + \pp_+ Z_1 \pp_+ \bar Z_1 + \pp_+ Z_2 \pp_+ \bar Z_2 \ ,
\ee
which sets
\be
(P^0)^2 = \frac{J_1 + 2 J_2}{\apr} \ .
\ee

In $D \geq 5$, spinning strings can carry
angular momenta $J_{1,2}$ in one or two planes, and the large-$J$ perturbation
theory is understood to keep these quantities in fixed ratio.
As described in \cite{PRL},
in a suitable Cartan decomposition, the angular momenta
are aligned with the ``3'' direction of the self-dual and antiself-dual
$SU(2)_\pm$ subgroups of the $SO(4)$ little group of $SO(D-1)$.
In $D\geq 5$, states can carry angular momenta in both planes
with angular-momentum quantum numbers
\be
J_\pm = \frac{1}{2}(J_1 \pm J_2) \ .
\ee
States are determined by minimizing the
energy over highest-weight vectors of $SU(2)\ll+ \times SU(2)\ll-$,
with total angular momenta $J\ll\pm$ and zero momentum in
the $\s^1$ direction. The free-field ground state in the open-string sector is unique
and can be expressed as
\be
|J_+,J_-;P\rangle_{\rm free} = \frac{1}{\sqrt{{\cal N}_{J_+,J_-}^{\rm
 (open)}}} \left(\a\ll{-1}^{Z_1}\a\ll{-2}^{Z_2} -
 \a\ll{-2}^{Z_1}\a\ll{-1}^{Z_2}\right)^{J_+-J_-}
 \left(\a_{-1}^{Z_1}\right)^{2J_-}|0;P\rangle_{\rm free} \ .
\ee
The quantity ${\cal N}_{J_+,J_-}^{\rm(open)}$ is a normalization
constant, and the energy under the free-field Hamiltonian takes the
form
\be
E^{\rm (free)} = \apr P^2 + 3J_+ - J_- - \frac{D}{24} \ .
\ee

Starting with $\widehat{\inv}\ll{22}$ in eqn.~\rr{I22hatDef} above,
\be
\widehat \inv\ll{22} \equiv \inv\ll{22} - \frac{\inv\ll{12} \inv\ll{21}}{\inv\ll{11}} \ ,
\ee
and adopting notation consistent with \cite{Hellerman:2014cba}, we introduce a 
regulated version of the Liouville field:
\be
\phc \equiv -\frac{1}{4} \log\left(\inv\ll{11}^2 - L^2 \widehat \inv\ll{22} \right) \ .
\label{RegLiouvilleField}\ee
The operator $\widehat{\inv}\ll{22}$ is a Weyl tensor of weight four, 
so the object $\phc$ transforms as a scalar under worldsheet diffeomorphisms and as a
Liouville field under Weyl transformations of
the intrinsic metric:
\bbb
\phc \to \phc + \r \llsk {\rm under} \llsk
g\ll{\bullet\bullet} \to \expp{2\r} \cc g\ll{\bullet\bullet} \ .
\een{LiouvilleTransformation}
Therefore, the anomaly action evaluated on $\phc$ has precisely the same anomaly-canceling property as the
anomaly action evaluated with $L = 0$, which leads to the Polchinski-Strominger anomaly term. 
In terms of $\phc$, 
the regulated anomaly term can conveniently be expressed as \cite{Hellerman:2014cba}
\bbb
{\cal L}\ll{\rm anom} \equiv  \frac{\b}{2\pi}\left(-|\gg\phc|\sqd +
\phc\cc R\lp 2 \right) \ ,
\eee
where $R\lp 2$ is the Ricci scalar curvature of the two-dimensional intrinsic metric.

After gauge-fixing $g\ll{ab} \to \eta\ll{ab}$, the anomaly
term becomes
\be
{\cal L}_{\rm anom} = \frac{2\b}{\pi} \pp_+ \phc\, \pp_- \phc \ .
\label{GeneralAnomalyAction}\ee
The path integral with the addition of this term is fully gauge invariant and finite at the boundary.

We now turn to the evaluation of this term in the classical helical solution \rr{ClassicalHelical}.
For $p,q \leq 2$, the ground state profile of $\inv\ll{pq}$ is as follows:
\be
\inv\ll{11} &=& -2\apr\left(J_1 + 4 J_2 + 4 J_2 \cos(2\s_1)\right) \sin^2(\s_1) 
\nn\\
\inv\ll{12} &=& -\inv\ll{21} = \apr\left(J_1 + 8 J_2 \cos(2\s_1)\right) \sin(2\s_1)
\nn\\
\inv\ll{22} &=& \apr\left(J_1 \cos(2\s_1) + 8 J_2 \cos(4\s_1)\right) \ .
\label{InvariantsInHelicalSol}
\ee
Taking these together, we recover an explicit expression for the Weyl-covariant version of 
$\inv\ll{22}$ on the ground state solution
\be
\widehat \inv\ll{22} &=& \apr\left(J_1\cos(2\s_1) + 8 J_2\cos(4\s_1)
    -2\cos^2(\s_1)\frac{\left(J_1+8 J_2\cos^2(2\s_1)\right)^2}{J_1+4 J_2+4 J_2\cos(2\s_1)} \right) \ .
    \nn\\
\ee

Analysis of the PS anomaly contribution can thus be reduced to a straightforward
contour integral evaluated by residues, with the removal of a UV divergence at the endpoints of the interval.
Let us introduce the following change of variables:
\be
\s_1 = \frac{1}{2i}\log w\ , \qquad d\s_1 = \frac{1}{2i}\frac{dw}{w} \ .
\ee
We can infer the location of the poles of the PS integrand (as a function of $w$) by looking
at the denominator of the PS anomaly term.  
To do this, let us further define
\be
a \equiv \frac{J_2}{J_1}\ , \qquad  b \equiv  \frac{L}{J_1^{1/2}} \ ,
\ee
and write the integrated Lagrangian as
\be
\int {\cal L}_{\rm PS} \, d\s_1 & = & -i \frac{\b}{4\pi} \int  \frac{(w+1)^2 (w-1)^6 F_1^2}{w\left[ w+2a(w+1)^2 \right]^2 F_2^2}\, dw \ .
\ee
The functions $F_1$ and $F_2$ are complicated polynomials in $w$, with coefficients depending on $a$, $b$ and $\apr$.
Organizing the polynomial coefficients according to
\be
F_1 &=& \sum_{i=0}\uu 8 C_{1,i} w^i \ ,
\nn\\
F_2 &=& \sum_{i=0}\uu{10} C_{2,i} w^i \ ,
\ee
we have 
\be
C_{1,0} & = & 32 a^4 \apr   \nn\\
C_{1,1} & = & 8 a^3(7+24a)\apr        \nn\\
C_{1,2} & = & 4 a^2\left(a+4a(15+32 a)\right) \apr        \nn\\
C_{1,3} & = & 2a \left[ -2ab^2 + \left[5+4a\left(12 + a(57+104a)\right)\right]\apr\right]       \nn\\
C_{1,4} & = & -4a(1+6a)b^2 + \left[1+4a\left[3+2a\left(15+4a(17+30a)\right)\right]\right]\apr        \nn\\
C_{1,5} & = & 2a \left[ -2ab^2 + \left[5+4a\left(12 + a(57+104a)\right)\right]\apr\right]         \nn\\
C_{1,6} & = & 4 a^2\left(a+4a(15+32 a)\right) \apr        \nn\\
C_{1,7} & = &  8 a^3(7+24a)\apr         \nn\\
C_{1,8} & = &   32 a^4 \apr    \ ,
\ee
and
\be
C_{2,0} & = & 8 a^3 \apr   \nn\\
C_{2,1} & = & 4 a^2 (3+4 a) \apr\nn\\
C_{2,2} & = & 6 a (1-4 a^2) \apr\nn\\
C_{2,3} & = & -4 a b^2-(-1+4 a (3+4 a (3+4 a))) \apr\nn\\
C_{2,4} & = & 8 a (3+8 a) b^2+2 (-2-3 a+8 a^3) \apr\nn\\
C_{2,5} & = & 4 (1+6 a+32 a^2) b^2+6 (1+2 a) (1+2 a+8 a^2) \apr\nn\\
C_{2,6} & = & 8 a (3+8 a) b^2+2 (-2-3 a+8 a^3) \apr\nn\\
C_{2,7} & = & -4 a b^2-(-1+4 a (3+4 a (3+4 a))) \apr\nn\\
C_{2,8} & = & 6 a (1-4 a^2) \apr\nn\\
C_{2,9} & = & 4 a^2 (3+4 a) \apr\nn\\
C_{2,10} & = & 8 a^3 \apr \ .
\ee

The analysis of the contour integral can then be organized as follows.
Poles (single or multiple) of the integrand can be sorted into those that give a nonzero contribution
as $\sqrt{L}\to 0$, and those that give vanishing contributions as $\sqrt{L}\to 0$.
In the second category, we find poles that lie outside the unit circle in the $w$ plane
at sufficiently small $\sqrt{L}$, as well as poles that either disappear or exhibit vanishing residue
as $\sqrt{L} \to 0$.  In particular, any pole that approaches any point on the $w$ unit circle
other than the point $w=1$ as $\sqrt{L} \to 0$ fall into the latter category;  
as $\sqrt{L}\to 0$ we must recover the original unregulated integrand, which is smooth everywhere
on the unit circle except at the point $w=1$.

Among the poles that provide a nonzero contribution, we find
a set of poles that approach points interior to the unit circle as $\sqrt{L}\to 0$, 
and a set that approaches the point $w=1$ in the same limit.  
Contributions from the former set can be computed by setting $\sqrt{L}$ to zero at the outset,
identifying poles interior to the unit circle, and calculating the corresponding residues.
Contributions from the second set can be determined by making the change of variables
$w \to 1 + i \sqrt{L} v$ and examining the limit $\sqrt{L}\to 0$ for fixed $v$.
In this limit, the positions of the singularities approach fixed locations in the $v$ plane,
and the residues scale\footnote{This scaling comes from a contribution of $\sqrt{L}$, 
strictly from the transformed measure, and a contribution of $L^{-1}$ from the leading-order
scaling of the original integrand, modulo the measure.} as $L^{-1/2}$. 
That is, in the scaling limit $L\to 0$, the PS term in these variables takes the form
\bbb
\int{\cal L}\ll{\rm PS}\, d\s_1 =
 L_{\rm PS}^{\rm (finite)} + L_{\rm PS}^{\rm (divergent)} + O(L\uu{1/2}) \ ,
 \een{FinitePlusDivergentAnomalyTermDecomposition}
 where the divergent term comes entirely from the scaling limit of the integral near the cluster of poles near $w=1$:
 \bbb
L_{\rm PS}^{\rm (divergent)} \equiv \int \left(-\frac{\b\apr^2}{\pi\sqrt{L}} \frac{q^2 v^6}{(4+\apr q v^4)^2} \right)\, dv \ ,
\ee
and $q$ is the combination 
\be
q \equiv J\ll 1 + 8 J\ll 2\ .
\ee
The value of the divergent term is:
\bbb
L_{\rm PS}^{{\rm (divergent)}} = - {{3\b (q\apr)\uu{{1\over 4}}}\over{8 \sqrt{L}}} 
\eee
Let us emphasize here that the $L^{-1/2}$ divergences are strictly proportional 
to the term ${\cal O}_{\rm quark} = \binv\lp{22}^{1/ 4}$.  
(Indeed, according to the dressing structure and, correspondingly, 
the allowed spectrum of boundary operators in the effective theory, this is the only possibility.)  The
particular combination $J\ll 1 + 8 J\ll 2$ occurring inside the fourth root agrees nontrivially with 
$\binv\lp{22}$, which can be read, e.g., from the boundary value of $\inv\ll{11}$ in \rr{InvariantsInHelicalSol}.
In the helical solution, this operator scales as 
$\langle {\cal O}_{\rm quark}\rangle \propto (J_1 + 8J_2)^{1/ 4} = q^{1/ 4}$, so the 
divergence of the PS integrand thus appears as $\langle {\cal O}_{\rm quark}\rangle$ with 
a coefficient that diverges as $L\uu{-1/2}$.   There are also terms in the integrand of order
$\sqrt{L}^0$, but these turn out to be odd in $v$, and thus integrate to zero.

The finite terms come from poles interior to the unit circle in the $L\to 0$ limit, and the sum of their residues can be found by integrating
along a circle enclosing all the poles away from $w\sim 1$, but excluding the poles near $w=1$.  In the $L=0$ expression, the interior
singularities comprise a single pole at the origin, and a double pole at 
\be
w\uu{\rm int}\lp{*} \equiv  {1\over{4 J\ll 2}} \cc \big ( J\ll
 1\uu{1/2} \sqrt{J\ll 1 + 8 J\ll 2} -  J\ll 1 - 4 J\ll 2\big )\ ,
\ee
which is always real and lies between $0$ and $-1$.

Altogether, the integral decomposes into contributions that are manifestly regulator-independent 
(i.e., those that approach interior points to the unit circle as $\sqrt{L}\to 0$), and 
contributions from a purely local UV divergence (i.e., the sum of contributions from poles
that approach $w=1$ as $\sqrt{L}\to 0$).  The contribution from the pole at the origin is
\bbb
\frac{2\b}{\pi}\oint \ll{w\to 0} \pp_+\phc\, \pp_-\phc \, d\s  = 2\b\ ,
\een{OriginPoleContrib}
and the contribution from the double pole at $w\uu{\rm int}\lp{*}$ is
\bbb
\frac{2\b}{\pi}\oint \ll{w\to w\uu{\rm int}\lp{*}} \pp_+\phc\, \pp_-\phc \, d\s  =  - {\b\over 2}\cc \frac{3J_1 + 4J_2}{\sqrt{J_1(J_1+8J_2)}}\ .
\eee
Thus, the UV-finite part of the PS anomaly term
\rr{FinitePlusDivergentAnomalyTermDecomposition} evaluates to 
\be
L_{\rm PS}^{{\rm (finite)}} = \frac{2\b}{\pi}\int \pp_+\phc\, \pp_-\phc \, d\s  = \frac{\b}{2}\biggl(4 - \frac{3J_1 + 4J_2}{\sqrt{J_1(J_1+8J_2)}}\biggr) \ ,
\ee
while the divergent piece can be removed with a boundary
counterterm proportional to the quark mass operator.  
As described in \cite{PRL}, the first-order shift in the energy of the lowest
classical solution with fixed Noether charges is 
just the negative of the interaction Lagrangian for the unperturbed, 
zeroth-order helically symmetric solution.  Replacing $\b = (26-D)/12$ (see \cite{Polchinski:1991ax,PRL}), we recover 
the contribution to the open string mass-squared from the PS interaction:
\be
\Delta M_{\rm open}^2 = \frac{D-26}{24\,\apr} \biggl(4 - \frac{3J_1 + 4J_2}{\sqrt{J_1(J_1+8J_2)}}\biggr) \ .
\ee

\appendix{Properties of geodesics near the boundary}
\label{GeodDistBdyFunc}
In the \sreg analysis of Section \ref{FiniteSepReg}, we introduced a
near-boundary cutoff scheme by defining a strip to be excised from
the worldsheet along a set of points on the
$\s\ll 1 = \tilde\s$ locus separated from the boundary by a fixed geodesic
distance $\epsilon$.  Here we demonstrate in detail the
gauge-invariant characterization of this distance function. In
particular, we identify the longest\footnote{In Lorentzian signature.}
spacelike geodesic in the near-boundary region, extending from the boundary to the interior point
$\tilde\sigma$.  We can show that such a global
maximum must always exist by first anchoring a point in the bulk.  Given a point on the boundary, there
is always a geodesic of some kind from the anchor point to that
boundary point. For boundary points sufficiently far in the past or the future,
the geodesic will be future-oriented timelike or past-oriented timelike.  Between, it must necessarily be spacelike. 
The geodesic length varies
continuously across this region of the boundary, so it must assume a
global maximum.  (On the endpoints of the spacelike-separated region, 
the proper length goes to zero, so the maximum is never assumed at the endpoints.)
For the static geometry induced by the helical
solution, this geodesic is just the horizontal trajectory in the obvious flat
coordinates.

Let us now make this argument more concrete.
With the expansion of $\inv\ll{11}$ in the near-boundary region
\rr{GeneralI11Expansion}, 
we can characterize the form of the metric in this region as
\be
\widehat{ds^2} \equiv{{ds\sqd}\over{\binv\lp{22}}}  =  \s\ll 1^2(-d\s\ll 0^2 + d\s\ll 1^2) \ .
\ee
It is convenient to make the following change of variables:
\be
a \equiv \frac{\s\ll 1^2}{2} \ .
\ee
We can always parameterize sufficiently short geodesics as functions
of $\s\ll 1$, so, in turn, we define
\be
b \equiv \sqrt{2} \s\uu 0 \equiv h(a) \ .
\ee

Working up to an overall scaling of the metric, we have
\be
\widehat{ds^2} = da^2 - a\, db^2 = da^2 - a (h')^2 da^2 = (1-a(h')^2)\, da^2\ .
\ee
The arc length is then
\be
\hat{\ell} = \int \sqrt{1 - a (h')^2}\, da\ ,
\ee
such that the geodesic equation on this space is just
\be
0 = \pp_a\left( \frac{a h'}{\sqrt{1-a(h')^2}}  \right) \longrightarrow
\frac{a h'}{\sqrt{1-a (h')^2}} = K \ ,
\ee
where $K$ is a constant.  Solving for $(h')^2$, 
\be
(h')^2 = \frac{K^2}{a(a+K^2)} \ ,
\ee
the geodesic equation admits solutions of the form
\be
h(a) = {\rm const.} \pm 2 K \log\left(\sqrt{a} + \sqrt{a+K^2} \right)\ .
\ee
When $K$ is nonzero, at small $\s_1$ (correspondingly, at small $a$),
the solutions are of the form 
\be
h(a) \biggr|_{a~{\rm small}} \approx {\rm const.} + ({\rm linear~in~}\s\ll 1)  + \cdots
\ee
As we approach the boundary, solutions with vanishing $K$ are 
asymptotically purely spacelike and normal to the boundary.  

One concern might have been that the singularity at the boundary might
spoil this analysis, though it does not.  
The integrated geodesic length remains finite, for instance, as a
function of the natural conformal coordinate $\s\ll 1$.  

More directly, we can compute the arc length of these geodesics from the boundary to an
anchor point $a_0$ in the interior.  We obtain
\be
\hat{\ell} &=& \int_0^{a_0}\sqrt{1 - \frac{K^2}{a+K^2}}\, da 
\nn\\
&=& \sqrt{a_0\left(a_0+K^2\right)} + K^2 \log\left(\frac{K}{\sqrt{a_0}+\sqrt{a_0+K^2}}\right) \ .
\ee
It can be shown that the $K = 0$ class of solutions globally
maximizes the length function in the asymptotic near-boundary region.
In detail, the first derivative of $l$ with respect to $K$ is
\be
\frac{\pp \hat{\ell}}{\pp K} &=& 2K\left(\sqrt{\frac{a_0}{a_0+K^2}} 
	+ \log\left(\frac{K}{\sqrt{a_0} + \sqrt{a_0+K^2}}\right)\right)
	\nn\\
	&\approx& (2-\log(4a_0)+2\log(K))K -\frac{3K^3}{2 a_0} + O(K^5) \ ,
\ee
which vanishes in the limit $K\to 0$.  The second derivative
\be
\frac{\pp^2 \hat{\ell}}{\pp K^2} & = & 2\left(\frac{\sqrt{a_0}(2a_0+K^2)}{(a_0+K^2)^{3/2}} + \log K
	- \log\left(\sqrt{a_0}+\sqrt{a_0+K^2}\right) \right)
	\nn\\ 
	&\approx& 4 - \log(4a_0) + 2\log K - \frac{9K^2}{2a_0} + O(K^4)
\ee
is negative as $K\to 0$.

\section*{Acknowledgments}
\addcontentsline{toc}{section}{Acknowledgments}
The authors are deeply grateful to J.~Sonnenschein and O.~Aharony for discussions that were responsible for
refining the ideas presented herein and motivating the
derivation of the dressing rule in section \ref{Liouville}.  The work of SH is supported by
the World Premier International Research Center Initiative (WPI
Initiative), MEXT, Japan, and also supported in part by JSPS KAKENHI Grant Numbers JP22740153, JP26400242.  SH and IS are grateful to the Walter Burke Institute for
Theoretical Physics at Caltech for generous hospitality while this work was in progress.


\end{document}